\documentclass[11pt]{article}
\usepackage{amssymb}
\usepackage[space]{cite}
\pdfoutput=1
\usepackage[T1]{fontenc}
\usepackage{multirow}
\usepackage{hyperref}
\usepackage[pdftex]{color,graphicx}

\begin{document}

\author{A.~Bernotas$^{a}$ and V.~\v{S}imonis$^{b}$, \and $^{a}${\small Vilnius
University Faculty of Physics, Saul\.{e}tekio 9, LT-10222 Vilnius, Lithuania}
\and $^{b}${\small Vilnius University Institute of Theoretical Physics and
Astronomy,} \and {\small \ A. Go\v{s}tauto 12, LT-01108,\ Vilnius, Lithuania}}
\title{MAGNETIC MOMENTS OF HEAVY BARYONS IN THE BAG MODEL REEXAMINED}
\date{{\small Received {\today}}}
\maketitle

\begin{abstract}
Magnetic moments of $J=\frac{1}{2}$ and $J=\frac{3}{2}$ heavy baryons are
calculated in the bag model with center-of-mass motion corrections. For the $%
\mathrm{spin\,}\frac{1}{2}$ baryons containing three quarks of different
flavours the effect of hyperfine mixing is examined in detail. The
results of the work are compared with predictions obtained in various other
approaches and models.

\medskip


\end{abstract}

\section{Introduction}

In recent years the interest in properties of heavy baryons has
grown (for review see e.g. \cite{KR10}). A relevant source of information
about the internal structure of particles is their magnetic moments.
Magnetic moments of heavy baryons have been considered in various
approaches. Nevertheless, the most extensive evaluation of these quantities
remains the bag model calculations performed more than 30 years ago \cite
{BS80}. Sometimes these bag model predictions still may serve as a useful
guide, yet several drawbacks are also evident. Firstly, only the ratios
of magnetic moments to that of proton were presented. Absolute values in
analogy with the case of light baryons were expected to be somewhat too
small. Secondly, the list of ground state baryons used in Ref.~\cite{BS80}
was incomplete -- the states $\Xi _{bc}$, $\Xi _{bc}^{\prime }$, $\Omega
_{bc} $, $\Omega _{bc}^{\prime }$, $\Omega _{bcc}$, and $\Omega _{bbc}$ were
missing. Thirdly, the hyperfine mixing of $%
\Xi _{c}$, $\Xi _{c}^{\prime }$ states (as well as $\Xi _{b}$, $\Xi
_{b}^{\prime }$) was not taken into account, which, as was shown in \cite{FLNC81}, can change the
predictions for magnetic moments substantially. Finally, the parameters
used in calculations of magnetic moments were chosen rather arbitrary.
The bag radii of charmed and bottom baryons were taken from
different variants of the bag model (Ref.~\cite{JK76} and Ref.~\cite{P79}
respectively), and the mass of the charmed quark did not correspond to any
of them. Therefore, the predictions of Ref.~\cite{BS80} for baryons
containing charmed quarks cannot be treated as exact bag model result, but
rather as a (more or less) crude estimate. We think that the contemporary
update of the bag model predictions is necessary. In a recent paper \cite
{BS12} we have used an improved bag model to calculate the magnetic moments
of light baryons. It was shown that the bag model with the center-of-mass
motion (c.m.m.) corrections taken into account can provide sufficiently good
predictions for magnetic moments. Now we are going to apply this model to
calculate magnetic moments of $J=\frac{1}{2}$ and $J=\frac{3}{2}$ heavy
baryons.

The paper is structured as follows. In Sec. 2 we present briefly the model
and give expressions for the baryon magnetic moments. In Sec. 3 we discuss
quark model relations (sum rules) that connect magnetic moments of different
baryons. Section 4 is devoted completely to the state mixing problem. Results
of our calculations are presented in Sec. 5. The latter also serves for
discussion and contains concluding remarks.

\section{Preliminaries: bag model, magnetic moments}

In our previous paper \cite{BS12} two slightly different variants of the bag
model suitable to provide a correct scale of the baryon magnetic moments
were considered. In the present work we have chosen to deal with
a first variant of these two because of its simplicity and universality. Below
we will briefly outline the main features of this bag model variant.

The hadron bag energy depends on the bag radius $R$ and is given by
\begin{equation}
E=\frac{4\pi }{3}BR^{3}+\frac{Z_{0}}{R}+\sum\limits_{i}\varepsilon
_{i}+\Delta E \, ,  \label{eq2.01a}
\end{equation}
where the four terms in the expression are: the bag volume energy, the Casimir
energy, the sum of single-particle eigenenergies, and the quark-quark
interaction energy (for more details see \cite{BS04}).

The interaction energy is defined to the first order in the effective
(running) coupling constant 
\begin{equation}
\alpha _{\mathrm{c}}(R)=\frac{2\pi }{9\ln (A+R_{0}/R)}\,\,,  \label{eq2.02a}
\end{equation}
where $A$ and $R_{0}$ are model parameters. Up and down quarks are assumed
to be massless. For heavier (strange, charmed, and bottom) quarks we use the
running mass defined as 
\begin{equation}
\overline{m}_{f}(R)=\widetilde{m}_{f}+\alpha _{\mathrm{c}}(R)\cdot \delta
_{f}\,,  \label{eq2.03a}
\end{equation}
where $\widetilde{m}_{f}$ and $\delta _{f}$ are additional flavour-dependent
parameters of the model.

The bag radius $R_{\mathrm{B}}$ of an individual hadron is obtained by minimizing (\ref
{eq2.01a}) with respect to $R$. The mass of the hadron is related to the
corresponding bag energy via expression
\begin{equation}
M^{2}=E^{2}-P^{2},  \label{eq2.04a}
\end{equation}
where the effective momentum square $P^{2}$ is defined as
\begin{equation}
P^{2}=\gamma \sum\limits_{i}p_{i}^{2}\,.  \label{eq2.05a}
\end{equation}
Here $p_{i}$ are the momenta of individual quarks, and the c.m.m.
parameter $\gamma $ is to be determined in a fitting procedure.

The c.m.m. corrected magnetic moments are given by the Halprin and Kerman \cite
{HK82} relation
\begin{equation}
\mu =\frac{E}{M}\ \mu ^{0}.  \label{eq2.06a}
\end{equation}

Altogether we have eleven free parameters in the model. These are: the bag
constant $B$, the Casimir energy parameter $Z_{0}$, the c.m.m. parameter $\gamma $,
two parameters from the definition of the running coupling constant ($A$ and 
$R_{0}$), and six parameters necessary to define the running mass functions
for the strange, charmed, and bottom quarks ($\widetilde{m}_{s}$, $\delta
_{s}$; $\widetilde{m}_{c}$, $\delta _{c}$; $\widetilde{m}_{b}$, $\delta _{b}$%
). The values of $B$, $Z_{0}$, $\gamma $, $A$, $R_{0}$, $\widetilde{m}_{s}$, 
$\delta _{s}$ were determined in Ref.~\cite{BS12}. They are: $B=7.468\cdot
10^{-4}~\mathrm{GeV}^{4}$, $Z_{0}=0.22$, $\gamma =2.153$, $A=0.6514$, $%
R_{0}=4.528~\mathrm{GeV}^{-1}$, $\widetilde{m}_{s}=0.262~\mathrm{GeV}$, $%
\delta _{s}=0.083$~$\mathrm{GeV}$. The numerical values of the remaining
four parameters $\widetilde{m}_{c}$, $\delta _{c}$, $\widetilde{m}_{b}$ and $%
\delta _{b}$ were obtained in the present work from the fit to the corresponding 
masses of $J/\psi $, $\Lambda _{c}$, $%
\Upsilon%
$, and $\Lambda _{b}$. They are: $\widetilde{m}_{c}=1.458~\mathrm{GeV}$, $%
\delta _{c}=0.089~\mathrm{GeV}$, $\widetilde{m}_{b}=4.721~\mathrm{GeV}$, and 
$\delta _{b}=0.079~\mathrm{GeV}$. As we see, the numerical values of
parameter $\delta _{f}$ for all three flavours are similar. So, in
principle, we could even reduce the number of free parameters of the model
and use one average value (e.g., $\bar{\delta}=0.084~\mathrm{GeV}$) for
strange, charmed, and bottom quarks.

The wave functions of baryons can be constructed by coupling the spins
of the two first quarks to an intermediate spin $S$ and then adding the
third one to form the baryon with the resulting spin $J$. Proceeding in such
manner one can construct the so-called antisymmetric (with respect to the
interchange of $q_{1}$ and $q_{2}$) 
\begin{equation}
\left| \lbrack q_{1}q_{2}]q_{3}\right\rangle =\left|
(q_{1}q_{2})^{S=0}q_{3}\right\rangle  \label{eq2.07a}
\end{equation}
and symmetric
\begin{equation}
\left| \{q_{1}q_{2}\}q_{3}\right\rangle =\left|
(q_{1}q_{2})^{S=1}q_{3}\right\rangle  \label{eq2.08a}
\end{equation}
states. For identically flavoured quarks $q_{1}$ and $q_{2}$ only the
symmetric states are allowed.

The valence quark contribution to the magnetic moments of baryons is given
by \cite{FLNC81,JS77}
\begin{equation}
S=0,\,J=\frac{1}{2}\,;\quad \mu ([q_{1}q_{2}]q_{3})=\mu _{3}\,,  \label{eq2.09a}
\end{equation}
\begin{equation}
S=1,\,J=\frac{1}{2}\,;\quad \mu (\{q_{1}q_{2}\}q_{3})=\frac{1}{3}(2\mu _{1}+2\mu
_{2}-\mu _{3})\,,  \label{eq2.10a}
\end{equation}
and 
\begin{equation}
S=1,\,J=\frac{3}{2}\,;\quad \mu (q_{1}q_{2}q_{3})=\mu _{1}+\mu _{2}+\mu _{3}\,,
\label{eq2.11a}
\end{equation}
where $\mu _{i}$ represents the magnetic moments of individual quarks. For
the transition moments we have 
\begin{equation}
\mu ^{\mathrm{tr}}(q_{1}q_{2}q_{3})=\mu (\{q_{1}q_{2}\}q_{3}\rightarrow
[q_{1}q_{2}]q_{3})=\frac{1}{\sqrt{3}}(\mu _{2}-\mu _{1})\,.  \label{eq2.12a}
\end{equation}
In the bag model magnetic moments of confined quarks are given by the
formula 
\begin{equation}
\mu _{i}=q_{i}\ \bar{\mu}_{i}\ ,  \label{eq2.13a}
\end{equation}
where $q_{i}$ is the electric charge of the quark, and reduced (charge-%
independent) quark magnetic moments $\bar{\mu}_{i}$ depend on the quark mass 
$m_{i}$, energy $\varepsilon _{i}$, and bag radius of baryon $R_{\mathrm{B}}$
(see \cite{DJJK75}): 
\begin{equation}
\bar{\mu}_{i}=\frac{4\varepsilon _{i}R_{\mathrm{B}}+2m_{i}R_{\mathrm{B}}-3}{%
2(\varepsilon _{i}R_{\mathrm{B}}-1)\varepsilon _{i}R_{\mathrm{B}}+m_{i}R_{%
\mathrm{B}}}\ \frac{R_{\mathrm{B}}}{6}\ .  \label{eq2.14a}
\end{equation}

The expression (\ref{eq2.12a}) changes its sign under the interchange of the
first two quarks in the wave functions (\ref{eq2.07a}), (\ref{eq2.08a}). This 
is important in the sense that relative signs of transition moments are
observable in an interference between decays to the same state. Therefore the
same quark ordering must be used consistently for each state of the same
quarks. Of course, we are free to choose the relative phase of the wave
function, but only once. Our phase conventions are the same as 
in Ref.~\cite{FLNC81} and differ from those adopted in e.g. Ref.~%
\cite{JS77}.

For the baryons containing three quarks of different flavours the
intermediate spins, in general, are not good quantum numbers because the
colour-magnetic interaction causes the mixing of the states (\ref{eq2.07a}) and 
(\ref{eq2.08a}). This problem will be discussed in detail in Sec.~4.

Using Eqs.~(\ref{eq2.09a})--(\ref{eq2.12a}) one can write down the explicit
expressions for the magnetic moments of all ground state heavy baryons.
Sometimes it is useful to have such expressions at hand. For convenience we
present them in the Appendix.

\section{Quark model relations and sum rules}

With some additional assumptions a plenty of quark model relations between
magnetic moments of various baryons can be obtained. The main assumption is
that quark magnetic moments in various baryons are the same. This is true
for the naive nonrelativistic quark model. In Ref.~\cite{JS77} this
assumption was used to obtain the relations connecting magnetic moments of
differently flavoured baryons (charmed and light, for example). Our opinion
is that there must be no illusions about the predictive power of such
relations.

In the bag model the quark magnetic moments can differ when passing from
baryon to baryon. Therefore the before mentioned assumption may be treated
as sufficiently accurate approximation only for baryons with the same (or
very similar) quark content. In other cases one should handle it with care.

From the expressions presented in the columns 3 of Tables~\ref{tA.1} and 
\ref{tA.3} (in the Appendix) we readily obtain the relations for magnetic moments of
charmed baryons (for simplicity we use shorthand notations in which magnetic
moments of particles are replaced by the symbols of these particles): 
\begin{equation}
\left. 
\begin{array}{c}
\Sigma _{c}^{++}+\Sigma _{c}^{0}=2\Sigma _{c}^{+}\,, \\[1ex] 
\Sigma _{c}^{*\,++}+\Sigma _{c}^{*\,0}=2\Sigma _{c}^{*\,+}\,,
\end{array}
\right.  \label{eq2.15a}
\end{equation}
\begin{equation}
\left. 
\begin{array}{c}
\Lambda _{c}^{+}+\Sigma _{c}^{+}=\frac{2}{3}\Sigma _{c}^{*\,+}\,, \\[1ex] 
\Xi _{c}^{0}+\Xi _{c}^{\prime \,0}=\frac{2}{3}\Xi _{c}^{*\,0}\,, \\[1ex] 
\Xi _{c}^{+}+\Xi _{c}^{\prime \,+}=\frac{2}{3}\Xi _{c}^{*\,+}\,,
\end{array}
\right.  \label{eq2.16a}
\end{equation}
\begin{equation}
\Sigma _{c}^{*\,++}+\Omega _{c}^{*\,0}=2\Xi _{c}^{*\,+}\,,  \label{eq2.17a}
\end{equation}
\begin{equation}
\left. 
\begin{array}{c}
4\Omega _{cc}^{+}+\Omega _{c}^{0}\approx 5\Lambda _{c}^{+}\,, \\[1ex] 
2\Omega _{cc}^{*\,+}-\Omega _{c}^{*\,0}\approx 3\Lambda _{c}^{+}\,,
\end{array}
\right.  \label{eq2.18a}
\end{equation}
\begin{equation}
\Omega _{ccc}^{++}\approx 3\Lambda _{c}^{+}\,,  \label{eq2.19a}
\end{equation}
\begin{equation}
\left. 
\begin{array}{c}
3[4\Omega _{c}^{0}+\Omega _{cc}^{+}]=5[2\Omega _{c}^{*\,0}-\Omega
_{cc}^{*\,+}]\,, \\[1ex] 
3[4\Omega _{cc}^{+}+\Omega _{c}^{0}]\approx 5[2\Omega _{cc}^{*\,+}-\Omega
_{c}^{*\,0}]\,.
\end{array}
\right.  \label{eq2.20a}
\end{equation}

In the naive nonrelativistic model there exist also relations $4\Omega
_{c}^{0}+\Omega _{cc}^{+}=5\Lambda $ and $2\Omega _{c}^{*\,0}-\Omega
_{cc}^{*\,+}=3\Lambda $, where $\Lambda $ represents the magnetic moment of
the strange baryon $\Lambda $. In the bag model these relations do not hold
anymore because magnetic moments of strange quarks entering light and heavy
baryons differ substantially. However, we can combine these equations to
obtain the first row of Eqs.~(\ref{eq2.20a}). In order to distinguish
between sufficiently accurate (in our model) and approximate relations we
use the symbol ``$=$'' in the cases when the accuracy of the relation is $%
\leq 3\%$, and the symbol ``$\approx $'' when the accuracy is in the range $%
(4-9)\%$. If the accuracy is worse, the relation is discarded.

An isospin symmetry leads to additional relations. For the magnetic moments
this symmetry means that one can set $\bar{\mu}_{u}=$ $\bar{\mu}_{d}$. Then
from columns 4 of Tables~\ref{tA.1} and \ref{tA.3} the following
relations can be deduced:
\begin{equation}
\left. 
\begin{array}{c}
\Sigma _{c}^{*\,0}=-3\Sigma _{c}^{+}\,, \\[1ex] 
\Sigma _{c}^{*\,++}=-3\Sigma _{c}^{0}\,, \\[1ex] 
2\Sigma _{c}^{+}+\Sigma _{c}^{0}=-\Lambda _{c}^{+}\,,
\end{array}
\right.  \label{eq2.21a}
\end{equation}
\begin{equation}
\Sigma _{c}^{*\,+}-\Sigma _{c}^{*\,0}=\Sigma _{c}^{*\,++}-\Sigma
_{c}^{*\,+}=\Xi _{c}^{*\,+}-\Xi _{c}^{*\,0}=\frac{3}{2}(\Sigma
_{c}^{+}-\Sigma _{c}^{0})\,,  \label{eq2.22a}
\end{equation}
\begin{equation}
\left. 
\begin{array}{c}
2\Xi _{cc}^{+}+\Xi _{cc}^{++}=\frac{4}{3}\Omega _{ccc}^{++}\,, \\[1ex] 
2\Xi _{cc}^{*\,+}+\Xi _{cc}^{*\,++}=2\Omega _{ccc}^{++}\,, \\[1ex] 
2\Sigma _{c}^{*\,+}+\Sigma _{c}^{*\,0}=3\Lambda _{c}^{+}\,, \\[1ex] 
\Xi _{cc}^{*\,++}-\Xi _{cc}^{*\,+}=3(\Xi _{cc}^{+}-\Xi _{cc}^{++})\,,
\end{array}
\right.  \label{eq2.23a}
\end{equation}
\begin{equation}
\Xi _{cc}^{*\,+}\approx \frac{3}{2}\Xi _{cc}^{++}\,,  \label{eq2.24a}
\end{equation}
\begin{equation}
\left. 
\begin{array}{c}
\Xi _{cc}^{*\,++}\approx 2\Sigma _{c}^{*\,+}\,, \\[1ex] 
\Xi _{cc}^{*\,++}-\Xi _{cc}^{*\,+}\approx \Xi _{c}^{*\,+}-\Xi _{c}^{*\,0}\,.
\end{array}
\right.  \label{eq2.25a}
\end{equation}

In the bottom sector the analogy of Eqs.~(\ref{eq2.15a})--(\ref{eq2.20a}) is
\begin{equation}
\left. 
\begin{array}{c}
\Sigma _{b}^{+}+\Sigma _{b}^{-}=2\Sigma _{b}^{0}\,, \\[1ex] 
\Sigma _{b}^{*\,+}+\Sigma _{b}^{*\,-}=2\Sigma _{b}^{*\,0},
\end{array}
\right.  \label{eq2.26a}
\end{equation}
\begin{equation}
\left. 
\begin{array}{c}
\Lambda _{b}^{0}+\Sigma _{b}^{0}=\frac{2}{3}\Sigma _{b}^{*\,0}\,, \\[1ex] 
\Xi _{b}^{-}+\Xi _{b}^{\prime \,-}=\frac{2}{3}\Xi _{b}^{*\,-}\,, \\[1ex] 
\Xi _{b}^{0}+\Xi _{b}^{\prime \,0}=\frac{2}{3}\Xi _{b}^{*\,0}\,, \\[1ex] 
\Xi _{bc}^{0}+\Xi _{bc}^{\prime \,0}=\frac{2}{3}\Xi _{bc}^{*\,0}\,, \\[1ex] 
\Xi _{bc}^{+}+\Xi _{bc}^{\prime \,+}=\frac{2}{3}\Xi _{bc}^{*\,+}\,, \\[1ex] 
\Omega _{bc}^{0}+\Omega _{bc}^{\prime \,0}=\frac{2}{3}\Omega _{bc}^{*\,0}\,,
\end{array}
\right.  \label{eq2.27a}
\end{equation}
\begin{equation}
\Sigma _{b}^{*\,+}+\Omega _{b}^{*\,-}=2\Xi _{b}^{*\,0}\,,  \label{eq2.28a}
\end{equation}
\begin{equation}
\Omega _{bbb}^{-}=3\Lambda _{b}^{0}\,.  \label{eq2.29a}
\end{equation}
\begin{equation}
\left. 
\begin{array}{c}
4\Omega _{bbc}^{0}+\Omega _{bcc}^{+}\approx 5\Lambda _{b}^{0}\,, \\[1ex] 
2\Omega _{bcc}^{*\,+}-\Omega _{bbc}^{*\,0}\approx \Omega _{ccc}^{++}\,, \\[1ex] 
2\Omega _{bbc}^{*\,0}-\Omega _{bcc}^{*\,+}\approx \Omega _{bbb}^{-}\,,
\end{array}
\right.  \label{eq2.30a}
\end{equation}
\begin{equation}
\left. 
\begin{array}{c}
3[4\Omega _{bbc}^{0}+\Omega _{bcc}^{+}]=5[2\Omega _{bbc}^{*\,0}-\Omega
_{bcc}^{*\,+}]\,, \\[1ex] 
3[4\Omega _{bcc}^{+}+\Omega _{bbc}^{0}]=5[2\Omega _{bcc}^{*\,+}-\Omega
_{bbc}^{*\,0}]\,, \\[1ex] 
3[4\Omega _{b}^{-}+\Omega _{bb}^{-}]=5[2\Omega _{b}^{*\,-}-\Omega
_{bb}^{*\,-}]\,.
\end{array}
\right.  \label{eq2.31a}
\end{equation}

In this case we have a problem with the quark model relations $4\Omega
_{bb}^{-}+\Omega _{b}^{-}=5\Lambda _{b}^{0}$ and $2\Omega
_{bb}^{*\,-}-\Omega _{b}^{*\,-}=3\Lambda _{b}^{0}$. In the bag model both of
them fail badly, and so does their combination $3[4\Omega _{bb}^{-}+\Omega
_{b}^{-}]=5[2\Omega _{bb}^{*\,-}-\Omega _{b}^{*\,-}]$. The culprit is the
strange quarks. The difference of their magnetic moments in the $\Omega
_{bb}^{-}$ and $\Omega _{b}^{-}$ baryons is comparable with the magnetic
moment of the bottom quark. As a consequence, the sufficiently accurate
value of the latter cannot be extracted from the above relations, and we
are forced to exclude them from our list.

The isospin symmetry now leads to the relations 
\begin{equation}
\left. 
\begin{array}{c}
\Sigma _{b}^{*\,-}=-3\Sigma _{b}^{0}\,, \\[1ex] 
\Sigma _{b}^{*\,+}=-3\Sigma _{b}^{-}\,, \\[1ex] 
\Xi _{bb}^{*\,-}=\frac{3}{2}\Xi _{bb}^{0}\,,
\end{array}
\right.  \label{eq2.32a}
\end{equation}
\begin{equation}
\Sigma _{b}^{*\,+}-\Sigma _{b}^{*\,0}=\Sigma _{b}^{*\,0}-\Sigma
_{b}^{*\,-}=\Xi _{b}^{*\,0}-\Xi _{b}^{*\,-}\,,  \label{eq2.33a}
\end{equation}
\begin{equation}
\left. 
\begin{array}{c}
\Xi _{bb}^{*\,0}-\Xi _{bb}^{*\,-}=3(\Xi _{bb}^{-}-\Xi _{bb}^{0})\,, \\[1ex] 
2\Xi _{bb}^{*\,-}+\Xi _{bb}^{*\,0}=2\Omega _{bbb}^{-}\,, \\[1ex] 
2\Xi _{bb}^{-}+\Xi _{bb}^{0}=4\Lambda _{b}^{0}\,,
\end{array}
\right.  \label{eq2.34a}
\end{equation}
\begin{equation}
\left. 
\begin{array}{c}
\Xi _{bb}^{*\,0}\approx 2\Sigma _{b}^{*\,0}\,, \\[1ex] 
\Xi _{bc}^{*\,+}-\Xi _{bc}^{*\,0}\approx \Xi _{b}^{*\,0}-\Xi _{b}^{*\,-}\,.
\end{array}
\right.  \label{eq2.35a}
\end{equation}

The states of the type $B,B^{\prime }$ enter the relations above only in the
combination $(B+B^{\prime })$. The reason is the state mixing which can
cause sizable shifts of unmixed quantities while leaving, however, the
combination $\mu (B)+\mu (B^{\prime })$ invariant. There are several states
the mixing of which is sufficiently small. First of all, such are two pairs of
states $\Lambda _{c}^{+},$ $\Sigma _{c}^{+}$ and $\Lambda _{b}^{0},$ $\Sigma
_{b}^{0}$, for which in the case of exact isospin symmetry there is not any
mixing at all. A careful analysis \cite{FLNC81} shows that this assumption
is valid to high degree also for real physical states. Therefore the mixing
between $\Lambda _{Q}$ and $\Sigma _{Q}$ states (where $Q$ denotes heavy
quark) can be safely ignored. Explicit calculations show that in the bag
model the mixing of $\Xi _{c}^{0}$, $\Xi _{c}^{\prime \,0}$ (as well as $\Xi
_{b}^{-}$, $\Xi _{b}^{\prime \,-}$) is not large enough to change the
unmixed magnetic moments substantially. Therefore we can add two more
relations to our collection: 
\begin{equation}
\left. 
\begin{array}{c}
\Xi _{c}^{0}\approx \Lambda _{c}^{+}\,, \\[1ex] 
\Xi _{b}^{-}\approx \Lambda _{b}^{0}\,.
\end{array}
\right.  \label{eq2.36a}
\end{equation}

That is all that remains from the naive nonrelativistic result 
\begin{eqnarray}
& \mu (\Lambda _{c}^{+})=\mu (\Xi _{c}^{0}) = \mu (\Xi _{c}^{+})\,, \nonumber
\nonumber \\[2ex]
& \mu (\Lambda _{b}^{0})=\mu (\Xi _{b}^{0})=\mu (\Xi _{b}^{-})=\mu (\Xi
_{bc}^{0})=\mu (\Xi _{bc}^{+})=\mu (\Omega _{bc}^{0})\,. \nonumber
\end{eqnarray}

One may wonder why one member of the isospin doublet (i.e. $\Xi _{c}^{+}$)
undergoes substantial changes, but another ($\Xi _{c}^{0}$) does not. The
reason is a very large $\Xi _{c}^{\prime \,+}\rightarrow \Xi _{c}^{+}$
transition magnetic moment, larger by an order of magnitude than the
corresponding $\Xi _{c}^{\prime \,0}\rightarrow \Xi _{c}^{0}$ moment. The
same is also true for the isospin doublet $\Xi _{b}^{-},\Xi _{b}^{0}$.

A reasonable question is if there could be any profit from all these nice
relations and sum rules. Of course, they can be used as a tool for the extra
check of the results obtained in calculations. They may help one to gain
some feeling (plausibly oversimplified) about the possible values of
magnetic moments under consideration and reveal some regularities as well. It
is not clear in general if many of them would survive in other approaches,
especially the more elaborated ones with various corrections included. In the
heavy hadron chiral perturbation theory \cite{BSBG00}, for example, Eqs.~(%
\ref{eq2.15a}), (\ref{eq2.17a}), (\ref{eq2.22a}), (\ref{eq2.26a}), (\ref{eq2.28a}%
), (\ref{eq2.33a}) are valid.

In the end, we think that caution must be paid while trying to use a single 
magnetic moment of baryons $\Xi _{c}^{+}$, $\Xi _{c}^{\prime \,+}$, $\Xi
_{b}^{0}$, $\Xi _{b}^{\prime \,0}$, $\Xi _{bc}^{0}$, $\Xi _{bc}^{\prime \,0}$%
, $\Xi _{bc}^{+}$, $\Xi _{bc}^{\prime \,+}$, $\Omega _{bc}^{0}$, and $\Omega
_{bc}^{\prime \,0}$ in the quark model sum rules such as, for example, $%
\Sigma _{c}^{++}+\Omega _{c}^{0}=2\Xi _{c}^{\prime \,+}$ \cite{BSBG00}.
State mixing effect can spoil this relation. In our model its accuracy is
only about $25\%$. The usual relations containing these states are valid
only when the unmixed states are considered. In the case of physical states
only the invariant combination of the type $B+B^{\prime }$ makes sense.

\section{Wave function mixing}

Among the heavy baryons under consideration there are some containing three
quarks of different flavours. In this case additional complications arise
because the quark-quark hyperfine interaction is not diagonal in the basis
defined by the wave functions~(\ref{eq2.07a}), (\ref{eq2.08a}). The physical 
states are the linear combinations of these states 
\begin{equation}
\left. 
\begin{array}{c}
\left| B\right\rangle \,=C_{1}\left| [q_{1}q_{2}]q_{3}\right\rangle
+C_{2}\left| \{q_{1}q_{2}\}q_{3}\right\rangle \,, \\[1ex] 
\left| B^{\prime }\right\rangle =-C_{2}\left| [q_{1}q_{2}]q_{3}\right\rangle
+C_{1}\left| \{q_{1}q_{2}\}q_{3}\right\rangle \,.
\end{array}
\right.  \label{eq3.1a}
\end{equation}

We have already studied the impact of such state mixing on the masses of
heavy baryons in Ref.~\cite{BS08}. Extensive study was also performed in the
framework of nonrelativistic potential model \cite{RP08,RP09}, and it was shown
that the state mixing has significant implications for some aspects of
phenomenology of these states such as their semileptonic decay rates. In
Ref.~\cite{FLNC81} it was shown that this mixing can affect the values of
magnetic moments even when it is not sufficiently strong to induce
significant shifts of baryon masses.

\begin{table}
\centering%
\caption{Mixed and unmixed magnetic moments (in nuclear magnetons) of
$\Xi _{c},\Xi _{c}^{\prime \,}$, and $\Xi _{b},\Xi _{b}^{\prime \,}$ baryons. \label{t3.1}}
\begin{tabular}{cccccc}
\hline
\multirow{2}{*}{Wave function} & $C_1$ & Magnetic & \multirow{2}{*}{Wave function} & $C_1$ & Magnetic \\
 & $C_2$ & moments &  & $C_2$ & moments \\
 \hline \\[-9pt]
$\left\{ 
\begin{array}{c}
\left| \Xi _{c}^{0}\right\rangle \\ 
\left| \Xi _{c}^{\prime \,0}\right\rangle \\ 
\left| \Xi _{c}^{\prime \,0}\right\rangle \rightarrow \left| \Xi
_{c}^{0}\right\rangle
\end{array}
\right. $
 &  & $\left. 
\begin{array}{r}
0.421 \\ 
-0.914 \\ 
0.128
\end{array}
\right. $ & $\left\{ 
\begin{array}{c}
\left| \Xi _{c}^{+}\right\rangle \\ 
\left| \Xi _{c}^{\prime \,+}\right\rangle \\ 
\left| \Xi _{c}^{\prime \,+}\right\rangle \rightarrow \left| \Xi
_{c}^{+}\right\rangle
\end{array}
\right. $ &  & $\left. 
\begin{array}{r}
0.257 \\ 
0.591 \\ 
-1.043
\end{array}
\right. $ \\ 
&  &  &  &  &  \\ 
$\left\{ 
\begin{array}{c}
\left| \lbrack ds]c\right\rangle \\ 
\left| \{ds\}c\right\rangle \\ 
\left| \{ds\}c\right\rangle \rightarrow \left| [ds]c\right\rangle
\end{array}
\right. $ & $
\begin{array}{r}
\phantom{-}0.997 \\ 
0.073 \\ 
\cdots\ \ 
\end{array}
$ & $\left. 
\begin{array}{r}
0.412 \\ 
-0.905 \\ 
0.110
\end{array}
\right. $ & $\left\{ 
\begin{array}{c}
\left| \lbrack us]c\right\rangle \\ 
\left| \{us\}c\right\rangle \\ 
\left| \{us\}c\right\rangle \rightarrow \left| [us]c\right\rangle
\end{array}
\right. $ & $
\begin{array}{r}
\phantom{-}0.997 \\ 
0.073 \\ 
\cdots\ \ 
\end{array}
$ & $\left. 
\begin{array}{r}
0.412 \\ 
0.438 \\ 
-1.057
\end{array}
\right. $ \\ 
&  &  &  &  &  \\ 
$\left\{ 
\begin{array}{c}
\begin{array}{c}
\left| \lbrack cd]s\right\rangle \\ 
\left| \{cd\}s\right\rangle
\end{array}
\\ 
\left| \{cd\}s\right\rangle \rightarrow \left| [cd]s\right\rangle
\end{array}
\right. $ & $
\begin{array}{r}
-0.562 \\ 
0.827 \\ 
\cdots\ \ 
\end{array}
$ & $\left. 
\begin{array}{r}
-0.482 \\ 
-0.015 \\ 
-0.626
\end{array}
\right. $ & $\left\{ 
\begin{array}{c}
\begin{array}{c}
\left| \lbrack cu]s\right\rangle \\ 
\left| \{cu\}s\right\rangle
\end{array}
\\ 
\left| \{cu\}s\right\rangle \rightarrow \left| [cu]s\right\rangle
\end{array}
\right. $ & $
\begin{array}{r}
-0.562 \\ 
0.827 \\ 
\cdots\ \ 
\end{array}
$ & $\left. 
\begin{array}{r}
-0.482 \\ 
1.335 \\ 
0.541
\end{array}
\right. $ \\ 
&  &  &  &  &  \\ 
$\left\{ 
\begin{array}{c}
\left| \lbrack sc]d\right\rangle \\ 
\left| \{sc\}d\right\rangle \\ 
\left| \{sc\}d\right\rangle \rightarrow \left| [sc]d\right\rangle
\end{array}
\right. $ & $\left. 
\begin{array}{r}
-0.435 \\ 
-0.900 \\ 
\cdots\ \ 
\end{array}
\right. $ & $\left. 
\begin{array}{r}
-0.672 \\ 
0.177 \\ 
0.516
\end{array}
\right. $ & $\left\{ 
\begin{array}{c}
\left| \lbrack sc]u\right\rangle \\ 
\left| \{sc\}u\right\rangle \\ 
\left| \{sc\}u\right\rangle \rightarrow \left| [sc]u\right\rangle
\end{array}
\right. $ & $
\begin{array}{r}
-0.435 \\ 
-0.900 \\ 
\cdots\ \ 
\end{array}
$ & $\left. 
\begin{array}{r}
1.344 \\ 
-0.498 \\ 
0.516
\end{array}
\right. $ \\ 
&  &  &  &  &  \\ 
$\left\{ 
\begin{array}{c}
\left| \Xi _{b}^{-}\right\rangle \\ 
\left| \Xi _{b}^{\prime \,-}\right\rangle \\ 
\left| \Xi _{b}^{\prime \,-}\right\rangle \rightarrow \left| \Xi
_{b}^{-}\right\rangle
\end{array}
\right. $ &  & $
\begin{array}{r}
-0.063 \\ 
-0.660 \\ 
0.082
\end{array}
$ & $\left\{ 
\begin{array}{c}
\left| \Xi _{b}^{0}\right\rangle \\ 
\left| \Xi _{b}^{\prime \,0}\right\rangle \\ 
\left| \Xi _{b}^{\prime \,0}\right\rangle \rightarrow \left| \Xi
_{b}^{0}\right\rangle
\end{array}
\right. $ &  & $\left. 
\begin{array}{r}
-0.100 \\ 
0.556 \\ 
-0.917
\end{array}
\right. $ \\ 
&  &  &  &  &  \\ 
$\left\{ 
\begin{array}{c}
\left| \lbrack ds]b\right\rangle \\ 
\left| \{ds\}b\right\rangle \\ 
\left| \{ds\}b\right\rangle \rightarrow \left| [ds]b\right\rangle
\end{array}
\right. $ & $
\begin{array}{r}
\phantom{-}0.999 \\ 
0.018 \\ 
\cdots\ \ 
\end{array}
$ & $
\begin{array}{r}
-0.066 \\ 
-0.656 \\ 
0.093
\end{array}
$ & $\left\{ 
\begin{array}{c}
\left| \lbrack us]b\right\rangle \\ 
\left| \{us\}b\right\rangle \\ 
\left| \{us\}b\right\rangle \rightarrow \left| [us]b\right\rangle
\end{array}
\right. $ & $
\begin{array}{r}
\phantom{-}0.999 \\ 
0.018 \\ 
\cdots\ \ 
\end{array}
$ & $\left. 
\begin{array}{r}
-0.066 \\ 
0.522 \\ 
-0.928
\end{array}
\right. $ \\ 
&  &  &  &  &  \\ 
$\left\{ 
\begin{array}{c}
\begin{array}{c}
\left| \lbrack bd]s\right\rangle \\ 
\left| \{bd\}s\right\rangle
\end{array}
\\ 
\left| \{bd\}s\right\rangle \rightarrow \left| [bd]s\right\rangle
\end{array}
\right. $ & $
\begin{array}{r}
-0.516 \\ 
0.857 \\ 
\cdots\ \ 
\end{array}
$ & $\left. 
\begin{array}{r}
-0.428 \\ 
-0.294 \\ 
-0.302
\end{array}
\right. $ & $\left\{ 
\begin{array}{c}
\begin{array}{c}
\left| \lbrack bu]s\right\rangle \\ 
\left| \{bu\}s\right\rangle
\end{array}
\\ 
\left| \{bu\}s\right\rangle \rightarrow \left| [bu]s\right\rangle
\end{array}
\right. $ & $
\begin{array}{r}
-0.516 \\ 
0.857 \\ 
\cdots\ \ 
\end{array}
$ & $\left. 
\begin{array}{r}
-0.428 \\ 
0.885 \\ 
0.719
\end{array}
\right. $ \\ 
&  &  &  &  &  \\ 
$\left\{ 
\begin{array}{c}
\left| \lbrack sb]d\right\rangle \\ 
\left| \{sb\}d\right\rangle \\ 
\left| \{sb\}d\right\rangle \rightarrow \left| [sb]d\right\rangle
\end{array}
\right. $ & $
\begin{array}{r}
-0.484 \\ 
-0.875 \\ 
\cdots\ \ 
\end{array}
$ & $\left. 
\begin{array}{r}
-0.589 \\ 
-0.133 \\ 
0.209
\end{array}
\right. $ & $\left\{ 
\begin{array}{c}
\left| \lbrack sb]u\right\rangle \\ 
\left| \{sb\}u\right\rangle \\ 
\left| \{sb\}u\right\rangle \rightarrow \left| [sb]u\right\rangle
\end{array}
\right. $ & $
\begin{array}{r}
-0.484 \\ 
-0.875 \\ 
\cdots\ \ 
\end{array}
$ & $\left. 
\begin{array}{r}
1.179 \\ 
-0.723 \\ 
0.209
\end{array}
\right. $ \\ \hline
\end{tabular}
\end{table}%

\begin{table}
\centering%
\caption{ Mixed and unmixed magnetic moments (in nuclear magnetons) of
doubly heavy baryons $\Xi _{bc},\Xi _{bc}^{\prime \,}$, and $\Omega
_{bc},\Omega _{bc}^{\prime \,}$.\label{t3.2}} 
\begin{tabular}{cccccc}
\hline
\multirow{2}{*}{Wave function} & $C_1$ & Magnetic & \multirow{2}{*}{Wave function} & $C_1$ & Magnetic \\
 & $C_2$ & moments &  & $C_2$ & moments \\
\hline \\[-9pt]
$\left\{ 
\begin{array}{c}
\left| \Xi _{bc}^{0}\right\rangle \\ 
\left| \Xi _{bc}^{\prime \,0}\right\rangle \\ 
\left| \Xi _{bc}^{\prime \,0}\right\rangle \rightarrow \left| \Xi
_{bc}^{0}\right\rangle
\end{array}
\right. $ &  & $
\begin{array}{r}
0.068 \\ 
-0.236 \\ 
0.508
\end{array}
$ & $\left\{ 
\begin{array}{c}
\left| \Xi _{bc}^{+}\right\rangle \\ 
\left| \Xi _{bc}^{\prime \,+}\right\rangle \\ 
\left| \Xi _{bc}^{\prime \,+}\right\rangle \rightarrow \left| \Xi
_{bc}^{+}\right\rangle
\end{array}
\right. $ &  & $
\begin{array}{r}
-0.157 \\ 
1.093 \\ 
-0.277
\end{array}
$ \\ 
&  &  &  &  &  \\ 
$\left\{ 
\begin{array}{c}
\left| \lbrack dc]b\right\rangle \\ 
\left| \{dc\}b\right\rangle \\ 
\left| \{dc\}b\right\rangle \rightarrow \left| [dc]b\right\rangle
\end{array}
\right. $ & $
\begin{array}{r}
\phantom{-}0.992 \\ 
0.128 \\ 
\cdots\ \ 
\end{array}
$ & $
\begin{array}{r}
-0.066 \\ 
-0.102 \\ 
0.530
\end{array}
$ & $\left\{ 
\begin{array}{c}
\left| \lbrack uc]b\right\rangle \\ 
\left| \{uc\}b\right\rangle \\ 
\left| \{uc\}b\right\rangle \rightarrow \left| [uc]b\right\rangle
\end{array}
\right. $ & $
\begin{array}{r}
\phantom{-}0.992 \\ 
0.128 \\ 
\cdots\ \ 
\end{array}
$ & $
\begin{array}{r}
-0.066 \\ 
1.002 \\ 
-0.427
\end{array}
$ \\ 
&  &  &  &  &  \\ 
$\left\{ 
\begin{array}{c}
\begin{array}{c}
\left| \lbrack bd]c\right\rangle \\ 
\left| \{bd\}c\right\rangle
\end{array}
\\ 
\left| \{bd\}c\right\rangle \rightarrow \left| [bd]c\right\rangle
\end{array}
\right. $ & $
\begin{array}{r}
-0.607 \\ 
0.795 \\ 
\cdots\ \ 
\end{array}
$ & $
\begin{array}{r}
0.366 \\ 
-0.534 \\ 
-0.281
\end{array}
$ & $\left\{ 
\begin{array}{c}
\begin{array}{c}
\left| \lbrack bu]c\right\rangle \\ 
\left| \{bu\}c\right\rangle
\end{array}
\\ 
\left| \{bu\}c\right\rangle \rightarrow \left| [bu]c\right\rangle
\end{array}
\right. $ & $
\begin{array}{r}
-0.607 \\ 
0.795 \\ 
\cdots\ \ 
\end{array}
$ & $
\begin{array}{r}
\phantom{-}0.366 \\ 
0.571 \\ 
0.676
\end{array}
$ \\ 
&  &  &  &  &  \\ 
$\left\{ 
\begin{array}{c}
\left| \lbrack cb]d\right\rangle \\ 
\left| \{cb\}d\right\rangle \\ 
\left| \{cb\}d\right\rangle \rightarrow \left| [cb]d\right\rangle
\end{array}
\right. $ & $
\begin{array}{r}
-0.385 \\ 
-0.923 \\ 
\cdots\ \ 
\end{array}
$ & $
\begin{array}{r}
-0.552 \\ 
0.384 \\ 
-0.249
\end{array}
$ & $\left\{ 
\begin{array}{c}
\left| \lbrack cb]u\right\rangle \\ 
\left| \{cb\}u\right\rangle \\ 
\left| \{cb\}u\right\rangle \rightarrow \left| [cb]u\right\rangle
\end{array}
\right. $ & $
\begin{array}{r}
-0.385 \\ 
-0.923 \\ 
\cdots\ \ 
\end{array}
$ & $
\begin{array}{r}
1.105 \\ 
-0.168 \\ 
-0.249
\end{array}
$ \\ 
&  &  &  &  &  \\ 
$\left\{ 
\begin{array}{c}
\left| \Omega _{bc}^{0}\right\rangle \\ 
\left| \Omega _{bc}^{\prime \,0}\right\rangle \\ 
\left| \Omega _{bc}^{\prime \,0}\right\rangle \rightarrow \left| \Omega
_{bc}^{0}\right\rangle
\end{array}
\right. $ &  & $
\begin{array}{r}
0.034 \\ 
-0.106 \\ 
0.443
\end{array}
$ &  &  &  \\ 
&  &  &  &  &  \\ 
$\left\{ 
\begin{array}{c}
\left| \lbrack sc]b\right\rangle \\ 
\left| \{sc\}b\right\rangle \\ 
\left| \{sc\}b\right\rangle \rightarrow \left| [sc]b\right\rangle
\end{array}
\right. $ & $
\begin{array}{r}
\phantom{-}0.994 \\ 
0.112 \\ 
\cdots\ \ 
\end{array}
$ & $
\begin{array}{r}
-0.066 \\ 
-0.007 \\ 
0.447
\end{array}
$ &  &  &  \\ 
&  &  &  &  &  \\ 
$\left\{ 
\begin{array}{c}
\begin{array}{c}
\left| \lbrack bs]c\right\rangle \\ 
\left| \{bs\}c\right\rangle
\end{array}
\\ 
\left| \{bs\}c\right\rangle \rightarrow \left| [bs]c\right\rangle
\end{array}
\right. $ & $
\begin{array}{r}
-0.593 \\ 
0.805 \\ 
\cdots\ \ 
\end{array}
$ & $
\begin{array}{r}
0.366 \\ 
-0.440 \\ 
-0.198
\end{array}
$ &  &  &  \\ 
&  &  &  &  &  \\ 
$\left\{ 
\begin{array}{c}
\left| \lbrack cb]s\right\rangle \\ 
\left| \{cb\}s\right\rangle \\ 
\left| \{cb\}s\right\rangle \rightarrow \left| [cb]s\right\rangle
\end{array}
\right. $ & $
\begin{array}{r}
-0.400 \\ 
-0.916 \\ 
\cdots\ \ 
\end{array}
$ & $
\begin{array}{r}
-0.409 \\ 
0.336 \\ 
-0.249
\end{array}
$ &  &  &  \\ \hline
\end{tabular}
\end{table}%

When the state mixing is taken into account the mixed magnetic moments of
the baryons are given by
\begin{equation}
\left. 
\begin{array}{c}
\mu (B)\ =C_{1}^{2}\,\mu ([q_{1}q_{2}]q_{3})+C_{2}^{2}\,\mu
(\{q_{1}q_{2}\}q_{3})+2C_{1}C_{2}\,\mu ^{\mathrm{tr}}(q_{1}q_{2}q_{3})\,,  \\[1ex] 
\ \mu (B^{\prime })=C_{1}^{2}\,\mu (\{q_{1}q_{2}\}q_{3})+C_{2}^{2}\,\mu
([q_{1}q_{2}]q_{3})-2C_{1}C_{2}\,\mu ^{\mathrm{tr}}(q_{1}q_{2}q_{3})\,.\,
\end{array}
\right.  \label{eq3.2}
\end{equation}
The physical transition moment is now
\begin{equation}
\mu (B^{\prime }\rightarrow B)=(C_{1}^{2}\,-C_{2}^{2}\,)\,\mu
^{\mathrm{tr}}(q_{1}q_{2}q_{3})+C_{1}C_{2}[\,\mu (\{q_{1}q_{2}\}q_{3})-\mu
([q_{1}q_{2}]q_{3})]\,.  \label{eq3.3}
\end{equation}

Without mixing the results in most cases depend very strongly on the quark
ordering in the spin coupling scheme $[(q_{1}q_{2})^{S}q_{3}]^{J}$. With
mixing the quark ordering becomes irrelevant, and in every case the final
result is the same. Various authors use in their calculations different
quark arrangements (very often with no state mixing taken into account), and
therefore sometimes it is not obvious how to compare our results with
other ones. To make things clearer, we present in Tables~\ref{t3.1}, 
\ref{t3.2} the results of our calculations of heavy baryon magnetic moments
with intermediate data for all possible quark orderings. We think it is a good
pedagogical example too. The calculation procedure is almost the same as
in Ref.~\cite{BS08}. The only difference is the opposite sign of the
off-diagonal matrix element of the interaction energy. We have changed the
relative phase of the wave functions in order to have the same phase
conventions as in Ref.~\cite{FLNC81}. The quark arrangements used are: $%
(q_{1}q_{2})^{S}q_{3}$, $(q_{3}q_{1})^{S}q_{2}$, and $(q_{2}q_{3})^{S}q_{1}$%
, where in the first one the quarks are ordered from lightest to heaviest.
Note that in order to maintain the relative phases of wave functions
unchanged the second and the third schemes are obtained from the first one by even
permutation of particles. Coefficients $C_{1}$, $C_{2}$ in Tables~%
\ref{t3.1}, \ref{t3.2} define the expansion of the mixed state $\left|
B\right\rangle =C_{1}\left| [q_{1}q_{2}]q_{3}\right\rangle +C_{2}\left|
\{q_{1}q_{2}\}q_{3}\right\rangle $.

From the inspection of Tables~\ref{t3.1} and \ref{t3.2} it is clear that
unmixed magnetic moments are very sensitive to the quark ordering scheme,
and therefore, strictly speaking, for magnetic moments there is no good
ordering scheme. For the states $\Xi _{c}^{0},\Xi _{c}^{\prime \,0}$ and $%
\Xi _{b}^{-},\Xi _{b}^{\prime \,-}$ the basis with the heaviest
quark standing in the end \cite{FLNC81,BS08} still remains preferable, 
but for other states ($\Xi _{c}^{+},\Xi _{c}^{\prime \,+}$; $\Xi
_{b}^{0},\Xi _{b}^{\prime \,0}$; $\Xi _{bc}^{0},\Xi _{bc}^{\prime \,0}$; $%
\Xi _{bc}^{+},\Xi _{bc}^{\prime \,+}$; and $\Omega _{bc}^{0},\Omega
_{bc}^{\prime \,0}$) the full account of the state mixing effect is
necessary.

\section{Results and discussion}

Now we are in a position to give the c.m.m. corrected bag model predictions
for magnetic moments of ground state heavy baryons. The results are
presented in Tables~\ref{t4.1}--\ref{t4.5}. We also compare our
predictions with results obtained using various other approaches. These
are:

\begin{itemize}
\item  Simple nonrelativistic quark model (Nonrel) with state mixing in the
case of baryons containing three differently flavoured quarks. Predictions
for mixed magnetic moments are taken from Ref.~\cite{FLNC81}, all other ones were
calculated using input values given in \cite{FLNC81} ($\mu _{u}=-2\mu _{d}$, 
$\mu _{d}=-0.93\,\mu _{N}$, $\mu _{s}=-0.61\,\mu _{N}$, $\mu _{c}=0.39\,\mu
_{N}$, and $\mu _{b}=-0.06\,\mu _{N}$) and explicit expressions from
Tables~\ref{tA.1}--\ref{tA.3}. Here and further $\mu _{N}$ denotes the
nuclear magneton.

\item  Phenomenological relativistic quark model \cite{JR04}. Authors of
this work studied three forms of relativistic kinematics. For comparison we
have picked out the ``instant'' form.

\item  Quark model based on the Dirac equation with a confining power-law
potential \cite{BD83}.

\item  Relativistic three-quark model \cite{FGIK06}.

\item  Full nonrelativistic calculation using Faddeev formalism with AL1
potential \cite{S96}.

\item  Nonrelativistic variational ansatz with the same AL1 potential \cite
{AHNV07}. For $J=\frac{3}{2}$ baryons (Tables~\ref{t4.3} and \ref{t4.5}) we
group the predictions of Ref.~\cite{S96} for magnetic moments of $\Omega
_{ccc}^{++}$ and $\Omega _{bbb}^{-}$ with the results obtained in \cite
{AHNV07} for other baryons in one column.

\item  Nonrelativistic model with screening and effective quark mass \cite
{KDV05,DV09}. By the way, in this approach the screening effect spoils the quark
model relation (\ref{eq2.24a}).

\item  Nonrelativistic hypercentral model \cite{PRV08,PMV09}. Their states $%
\Xi _{Q}$ are evidently symmetric and should be renamed as $\Xi _{Q}^{\prime
}$.

\item  Chiral constituent quark model \cite{SDCG10}. Their prediction for
the magnetic moment of triply heavy baryon $\Omega _{ccc}^{++}$ is an order
of magnitude lower than predictions obtained in all other Refs. We guess
that authors of \cite{SDCG10} have used for valence contribution the value $%
0.165$\thinspace $\mu _{N}$ instead of $1.165\,\mu _{N}$. In Table~\ref
{t4.3} we have changed their value for magnetic moment of $\Omega _{ccc}^{++}
$ from $0.155\,\mu _{N}$ to $1.17\,\mu _{N}$ on our own responsibility.

\item  Chiral perturbation theory \cite{S94}.

\item  QCD spectral sum rules \cite{ZHY97}.

\item  Light cone QCD sum rules \cite{AOS02,AAO08,AAO09}.
\end{itemize}

\begin{table}
\centering%
\caption{ Magnetic moments (in nuclear magnetons) of $J=\frac{1}{2}$ charmed
baryons calculated in the bag model (Bag) and in other approaches as
described in the text.\label{t4.1}} 
\begin{tabular}{lcccccccc}
\hline
Baryons & Bag & Nonrel & \cite{JR04} & \cite{BD83} & \cite{FGIK06} & \cite
{KDV05} & \cite{SDCG10} & \cite{S94} \\ \hline
$\Lambda _{c}^{+}$ & \multicolumn{1}{r}{$0.411$} & \multicolumn{1}{r}{$0.39$}
& \multicolumn{1}{r}{$0.40$} & \multicolumn{1}{r}{$0.341$} & 
\multicolumn{1}{r}{$0.42$} & \multicolumn{1}{r}{$0.37$} & \multicolumn{1}{r}{%
$0.39$$\phantom{0}$} & \multicolumn{1}{r}{$0.37$} \\[2pt] 
$\Sigma _{c}^{0}$ & \multicolumn{1}{r}{$-1.043$} & \multicolumn{1}{r}{$-1.37$%
} & \multicolumn{1}{r}{$-1.38$} & \multicolumn{1}{r}{$-1.391$} & 
\multicolumn{1}{r}{$-1.04$} & \multicolumn{1}{r}{$-1.17$} & 
\multicolumn{1}{r}{$-1.60$$\phantom{0}$} & --- \\[2pt] 
$\Sigma _{c}^{+}$ & \multicolumn{1}{r}{$0.318$} & \multicolumn{1}{r}{$0.49$}
& \multicolumn{1}{r}{$0.49$} & \multicolumn{1}{r}{$0.525$} & 
\multicolumn{1}{r}{$0.36$} & \multicolumn{1}{r}{$0.63$} & \multicolumn{1}{r}{%
$0.30$$\phantom{0}$} & --- \\[2pt] 
$\Sigma _{c}^{++}$ & \multicolumn{1}{r}{$1.679$} & \multicolumn{1}{r}{$2.35$}
& \multicolumn{1}{r}{$2.36$} & \multicolumn{1}{r}{$2.44$$\phantom{0}$} & 
\multicolumn{1}{r}{$1.76$} & \multicolumn{1}{r}{$2.18$} & \multicolumn{1}{r}{%
$2.20$$\phantom{0}$} & --- \\[2pt] 
$\Xi _{c}^{0}$ & \multicolumn{1}{r}{$0.421$} & \multicolumn{1}{r}{$0.41$} & 
\multicolumn{1}{r}{$0.41$} & \multicolumn{1}{r}{$0.341$} & 
\multicolumn{1}{r}{$0.39$} & \multicolumn{1}{r}{$0.36$} & \multicolumn{1}{r}{%
$0.28$$\phantom{0}$} & \multicolumn{1}{r}{$0.32$} \\[2pt] 
$\Xi _{c}^{\prime \,0}$ & \multicolumn{1}{r}{$-0.914$} & \multicolumn{1}{r}{$%
-1.18$} & \multicolumn{1}{r}{$-1.12$} & \multicolumn{1}{r}{$-1.12$$\phantom{0}$} & 
\multicolumn{1}{r}{$-0.95$} & \multicolumn{1}{r}{$-0.93$} & 
\multicolumn{1}{r}{$-1.32$$\phantom{0}$} & --- \\[2pt] 
$\Xi _{c}^{+}$ & \multicolumn{1}{r}{$0.257$} & \multicolumn{1}{r}{$0.20$} & 
\multicolumn{1}{r}{$0.40$} & \multicolumn{1}{r}{$0.341$} & 
\multicolumn{1}{r}{$0.41$} & \multicolumn{1}{r}{$0.37$} & \multicolumn{1}{r}{%
$0.40$$\phantom{0}$} & \multicolumn{1}{r}{$0.42$} \\[2pt] 
$\Xi _{c}^{\prime \,+}$ & \multicolumn{1}{r}{$0.591$} & \multicolumn{1}{r}{$%
0.89$} & \multicolumn{1}{r}{$0.75$} & \multicolumn{1}{r}{$0.796$} & 
\multicolumn{1}{r}{$0.47$} & \multicolumn{1}{r}{$0.76$} & \multicolumn{1}{r}{%
$0.76$$\phantom{0}$} & --- \\[2pt] 
$\Omega _{c}^{0}$ & \multicolumn{1}{r}{$-0.774$} & \multicolumn{1}{r}{$-0.94$%
} & \multicolumn{1}{r}{$-0.86$} & \multicolumn{1}{r}{$-0.850$} & 
\multicolumn{1}{r}{$-0.85$} & \multicolumn{1}{r}{$-0.92$} & 
\multicolumn{1}{r}{$-0.90$$\phantom{0}$} & --- \\[2pt] 
$\Xi _{cc}^{+}$ & \multicolumn{1}{r}{$0.722$} & \multicolumn{1}{r}{$0.83$} & 
\multicolumn{1}{r}{$0.86$} & \multicolumn{1}{r}{$0.774$} & 
\multicolumn{1}{r}{$0.72$} & \multicolumn{1}{r}{$0.77$} & \multicolumn{1}{r}{%
$0.84$$\phantom{0}$} & --- \\[2pt] 
$\Xi _{cc}^{++}$ & \multicolumn{1}{r}{$0.114$} & \multicolumn{1}{r}{$-0.10$}
& \multicolumn{1}{r}{$-0.10$} & \multicolumn{1}{r}{$-0.184$} & 
\multicolumn{1}{r}{$0.13$} & \multicolumn{1}{r}{$-0.11$} & 
\multicolumn{1}{r}{$0.006$} & --- \\[2pt] 
$\Omega _{cc}^{+}$ & \multicolumn{1}{r}{$0.668$} & \multicolumn{1}{r}{$0.72$}
& \multicolumn{1}{r}{$0.72$} & \multicolumn{1}{r}{$0.639$} & 
\multicolumn{1}{r}{$0.67$} & \multicolumn{1}{r}{$0.70$} & \multicolumn{1}{r}{%
$0.70$$\phantom{0}$} & --- \\[2pt] 
$\left| \Sigma _{c}^{+}\rightarrow \Lambda _{c}^{+}\right| $ & 
\multicolumn{1}{r}{$1.182$} & \multicolumn{1}{r}{$1.61$} & --- & --- & --- & 
\multicolumn{1}{r}{$1.54$} & \multicolumn{1}{r}{$1.56$$\phantom{0}$} & --- \\[2pt] 
$\left| \Xi _{c}^{\prime \,0}\rightarrow \Xi _{c}^{0}\right| $ & 
\multicolumn{1}{r}{$0.013$} & \multicolumn{1}{r}{$0.08$} & --- & --- & --- & 
\multicolumn{1}{r}{$0.13$} & \multicolumn{1}{r}{$0.31$$\phantom{0}$} & --- \\[2pt] 
$\left| \Xi _{c}^{\prime \,+}\rightarrow \Xi _{c}^{+}\right| $ & 
\multicolumn{1}{r}{$1.043$} & \multicolumn{1}{r}{$1.40$} & --- & --- & --- & 
\multicolumn{1}{r}{$1.39$} & \multicolumn{1}{r}{$1.30$$\phantom{0}$} & --- \\ \hline
\end{tabular}
\end{table}%

\begin{table}
\centering%
\caption{Magnetic moments (in nuclear magnetons) of $J=\frac{1}{2}$ charmed
baryons -- continuation of Table 3.\label{t4.2}} 
\begin{tabular}{lcccccc}
\hline
Baryons & Bag & \cite{S96} & \cite{AHNV07} & \cite{PRV08} & \cite{ZHY97} & 
\cite{AAO08} \\ \hline
$\Lambda _{c}^{+}$ & \multicolumn{1}{r}{$0.411$} & \multicolumn{1}{r}{$0.341$%
} & --- & \multicolumn{1}{r}{$0.385$} & \multicolumn{1}{r}{$0.15\pm 0.05$} & 
\multicolumn{1}{r}{$0.40\pm 0.05$} \\[2pt] 
$\Sigma _{c}^{0}$ & \multicolumn{1}{r}{$-1.043$} & \multicolumn{1}{r}{$%
-1.435 $} & --- & \multicolumn{1}{r}{$-1.015$} & \multicolumn{1}{r}{$-1.6\pm
0.2$$\phantom{0}$} & --- \\[2pt] 
$\Sigma _{c}^{+}$ & \multicolumn{1}{r}{$0.318$} & \multicolumn{1}{r}{$0.548$}
& --- & \multicolumn{1}{r}{$0.501$} & \multicolumn{1}{r}{$0.6\pm 0.1$$\phantom{0}$} & ---
\\[2pt] 
$\Sigma _{c}^{++}$ & \multicolumn{1}{r}{$1.679$} & \multicolumn{1}{r}{$2.532$%
} & --- & \multicolumn{1}{r}{$2.279$} & \multicolumn{1}{r}{$2.1\pm 0.3$$\phantom{0}$} & 
--- \\[2pt] 
$\Xi _{c}^{0}$ & \multicolumn{1}{r}{$0.421$} & \multicolumn{1}{r}{$0.360$} & 
--- & --- & --- & \multicolumn{1}{r}{$0.35\pm 0.05$} \\[2pt] 
$\Xi _{c}^{\prime \,0}$ & \multicolumn{1}{r}{$-0.914$} & --- & --- & 
\multicolumn{1}{r}{$-0.966$} & --- & --- \\[2pt] 
$\Xi _{c}^{+}$ & \multicolumn{1}{r}{$0.257$} & \multicolumn{1}{r}{$0.211$} & 
--- & --- & --- & \multicolumn{1}{r}{$0.50\pm 0.05$} \\[2pt] 
$\Xi _{c}^{\prime \,+}$ & \multicolumn{1}{r}{$0.591$} & --- & --- & 
\multicolumn{1}{r}{$0.711$} & --- & --- \\[2pt] 
$\Omega _{c}^{0}$ & \multicolumn{1}{r}{$-0.774$} & \multicolumn{1}{r}{$%
-0.835 $} & --- & \multicolumn{1}{r}{$-0.960$} & --- & --- \\[2pt] 
$\Xi _{cc}^{+}$ & \multicolumn{1}{r}{$0.722$} & \multicolumn{1}{r}{$0.784$}
& \multicolumn{1}{r}{$0.785$} & \multicolumn{1}{r}{$0.860$} & --- & --- \\[2pt] 
$\Xi _{cc}^{++}$ & \multicolumn{1}{r}{$0.114$} & \multicolumn{1}{r}{$-0.206$}
& \multicolumn{1}{r}{$-0.208$} & \multicolumn{1}{r}{$-0.137$} & --- & --- \\[2pt] 
$\Omega _{cc}^{+}$ & \multicolumn{1}{r}{$0.668$} & \multicolumn{1}{r}{$0.635$%
} & \multicolumn{1}{r}{$0.635$} & \multicolumn{1}{r}{$0.785$} & --- & --- \\
\hline
\end{tabular}
\end{table}%

\begin{table}
\centering%
\caption{Magnetic moments (in nuclear magnetons) of $J=\frac{3}{2}$ charmed
baryons calculated in the bag model (Bag) and in other approaches as
described in the text.\label{t4.3}} 
\begin{tabular}{lccccccc}
\hline
Baryons & Bag & Nonrel & \cite{S96,AHNV07} & \cite{DV09} & \cite{PRV08,PMV09}
& \cite{SDCG10}$^*$ & \cite{AAO09} \\ \hline
$\Sigma _{c}^{*\,0}$ & \multicolumn{1}{r}{$-0.958$} & \multicolumn{1}{r}{$%
-1.47$} & --- & \multicolumn{1}{r}{$-1.18$$\phantom{0}$} & \multicolumn{1}{r}{$-0.850$} & 
\multicolumn{1}{r}{$-1.99$} & \multicolumn{1}{r}{$-0.81\pm 0.20$} \\[2pt] 
$\Sigma _{c}^{*\,+}$ & \multicolumn{1}{r}{$1.085$} & \multicolumn{1}{r}{$%
1.32 $} & --- & \multicolumn{1}{r}{$1.18$$\phantom{0}$} & \multicolumn{1}{r}{$1.256$} & 
\multicolumn{1}{r}{$0.97$} & \multicolumn{1}{r}{$2.00\pm 0.46$} \\[2pt] 
$\Sigma _{c}^{*\,++}$ & \multicolumn{1}{r}{$3.127$} & \multicolumn{1}{r}{$%
4.11$} & --- & \multicolumn{1}{r}{$3.63$$\phantom{0}$} & \multicolumn{1}{r}{$3.844$} & 
\multicolumn{1}{r}{$3.92$} & \multicolumn{1}{r}{$4.81\pm 1.22$} \\[2pt] 
$\Xi _{c}^{*\,0}$ & \multicolumn{1}{r}{$-0.746$} & \multicolumn{1}{r}{$-1.15$%
} & --- & \multicolumn{1}{r}{$-1.02$$\phantom{0}$} & \multicolumn{1}{r}{$-0.690$} & 
\multicolumn{1}{r}{$-1.49$} & \multicolumn{1}{r}{$-0.68\pm 0.18$} \\[2pt] 
$\Xi _{c}^{*\,+}$ & \multicolumn{1}{r}{$1.270$} & \multicolumn{1}{r}{$1.64$}
& --- & \multicolumn{1}{r}{$1.39$$\phantom{0}$} & \multicolumn{1}{r}{$1.517$} & 
\multicolumn{1}{r}{$1.59$} & \multicolumn{1}{r}{$1.68\pm 0.42$} \\[2pt] 
$\Omega _{c}^{*\,0}$ & \multicolumn{1}{r}{$-0.547$} & \multicolumn{1}{r}{$%
-0.83$} & --- & \multicolumn{1}{r}{$-0.84$$\phantom{0}$} & \multicolumn{1}{r}{$-0.867$} & 
\multicolumn{1}{r}{$-0.86$} & \multicolumn{1}{r}{$-0.62\pm 0.18$} \\[2pt] 
$\Xi _{cc}^{*\,+}$ & \multicolumn{1}{r}{$0.163$} & \multicolumn{1}{r}{$-0.15$%
} & \multicolumn{1}{r}{$-0.311$} & \multicolumn{1}{r}{$0.035$} & 
\multicolumn{1}{r}{$-0.168$} & \multicolumn{1}{r}{$-0.47$} & --- \\[2pt] 
$\Xi _{cc}^{*\,++}$ & \multicolumn{1}{r}{$2.001$} & \multicolumn{1}{r}{$2.64$%
} & \multicolumn{1}{r}{$2.670$} & \multicolumn{1}{r}{$2.52$$\phantom{0}$} & 
\multicolumn{1}{r}{$2.755$} & \multicolumn{1}{r}{$2.66$} & --- \\[2pt] 
$\Omega _{cc}^{*\,+}$ & \multicolumn{1}{r}{$0.332$} & \multicolumn{1}{r}{$%
0.17$} & \multicolumn{1}{r}{$0.139$} & \multicolumn{1}{r}{$0.21$$\phantom{0}$} & 
\multicolumn{1}{r}{$0.121$} & \multicolumn{1}{r}{$0.14$} & --- \\[2pt] 
$\Omega _{ccc}^{++}$ & \multicolumn{1}{r}{$1.138$} & \multicolumn{1}{r}{$%
1.17 $} & \multicolumn{1}{r}{$1.023$} & \multicolumn{1}{r}{$1.16$$\phantom{0}$} & 
\multicolumn{1}{r}{$1.189$} & \multicolumn{1}{r}{$1.17$} & --- \\ \hline
\multicolumn{8}{l}{$^*$~Value for $\Omega _{ccc}^{++}$ corrected as deduced and explained in the text.}
\end{tabular}
\end{table}%

\begin{table}
\centering%
\caption{Magnetic moments (in nuclear magnetons) of $J=\frac{1}{2}$ bottom
baryons calculated in the bag model (Bag) and in other approaches as
described in the text.\label{t4.4}} 
\begin{tabular}{lcccccccc}
\hline
Baryons & Bag & Nonrel & \cite{BD83} & \cite{FGIK06}$^{**}$ & \cite{S96} & \cite
{AHNV07} & \cite{PRV08,PMV09} & \cite{AOS02,AAO08} \\ \hline
$\Lambda _{b}^{0}$ & \multicolumn{1}{r}{$-0.066$} & \multicolumn{1}{r}{$%
-0.06 $} & --- & \multicolumn{1}{r}{$-0.06$} & \multicolumn{1}{r}{$-0.060$}
& --- & \multicolumn{1}{r}{$-0.064$} & \multicolumn{1}{r}{$-0.18\pm 0.05$$\phantom{0}$}
\\[2pt] 
$\Sigma _{b}^{-}$ & \multicolumn{1}{r}{$-0.778$} & \multicolumn{1}{r}{$-1.22$%
} & \multicolumn{1}{r}{$-1.256$} & \multicolumn{1}{r}{$-1.01$} & 
\multicolumn{1}{r}{$-1.305$} & --- & \multicolumn{1}{r}{$-1.047$} & --- \\[2pt] 
$\Sigma _{b}^{0}$ & \multicolumn{1}{r}{$0.422$} & \multicolumn{1}{r}{$0.64$}
& \multicolumn{1}{r}{$0.659$} & \multicolumn{1}{r}{$0.53$} & 
\multicolumn{1}{r}{$0.682$} & --- & \multicolumn{1}{r}{$0.592$} & --- \\[2pt] 
$\Sigma _{b}^{+}$ & \multicolumn{1}{r}{$1.622$} & \multicolumn{1}{r}{$2.50$}
& \multicolumn{1}{r}{$2.575$} & \multicolumn{1}{r}{$2.07$} & 
\multicolumn{1}{r}{$2.669$} & --- & \multicolumn{1}{r}{$2.229$} & --- \\[2pt] 
$\Xi _{b}^{-}$ & \multicolumn{1}{r}{$-0.063$} & \multicolumn{1}{r}{$-0.05$}
& --- & \multicolumn{1}{r}{$-0.06$} & \multicolumn{1}{r}{$-0.055$} & --- & 
--- & \multicolumn{1}{r}{$-0.08\pm 0.02$$\phantom{0}$} \\[2pt] 
$\Xi _{b}^{\prime \,-}$ & \multicolumn{1}{r}{$-0.660$} & \multicolumn{1}{r}{$%
-1.02$} & \multicolumn{1}{r}{$-0.985$} & \multicolumn{1}{r}{$-0.91$} & --- & 
--- & \multicolumn{1}{r}{$-0.902$} & --- \\[2pt] 
$\Xi _{b}^{0}$ & \multicolumn{1}{r}{$-0.100$} & \multicolumn{1}{r}{$-0.11$}
& --- & \multicolumn{1}{r}{$-0.06$} & \multicolumn{1}{r}{$-0.086$} & --- & 
--- & \multicolumn{1}{r}{$-0.045\pm 0.005$} \\[2pt] 
$\Xi _{b}^{\prime \,0}$ & \multicolumn{1}{r}{$0.556$} & \multicolumn{1}{r}{$%
0.90$} & \multicolumn{1}{r}{$0.930$} & \multicolumn{1}{r}{$0.66$} & --- & ---
& \multicolumn{1}{r}{$0.766$} & --- \\[2pt] 
$\Omega _{b}^{-}$ & \multicolumn{1}{r}{$-0.545$} & \multicolumn{1}{r}{$-0.79$%
} & \multicolumn{1}{r}{$-0.714$} & \multicolumn{1}{r}{$-0.82$} & 
\multicolumn{1}{r}{$-0.703$} & --- & \multicolumn{1}{r}{$-0.960$} & --- \\[2pt] 
$\Xi _{bc}^{0}$ & \multicolumn{1}{r}{$0.068$} & \multicolumn{1}{r}{$0.13$} & 
--- & \multicolumn{1}{r}{$0.42$} & \multicolumn{1}{r}{$0.058$} & 
\multicolumn{1}{r}{$0.518$} & \multicolumn{1}{r}{$0.477$} & --- \\[2pt] 
$\Xi _{bc}^{\prime \,0}$ & \multicolumn{1}{r}{$-0.236$} & \multicolumn{1}{r}{%
$-0.53$} & \multicolumn{1}{r}{$-0.390$} & \multicolumn{1}{r}{$-0.76$} & ---
& \multicolumn{1}{r}{$-0.993$} & --- & --- \\[2pt] 
$\Xi _{bc}^{+}$ & \multicolumn{1}{r}{$-0.157$} & \multicolumn{1}{r}{$-0.25$}
& --- & \multicolumn{1}{r}{$-0.12$} & \multicolumn{1}{r}{$-0.198$} & 
\multicolumn{1}{r}{$-0.475$} & \multicolumn{1}{r}{$-0.400$} & --- \\[2pt] 
$\Xi _{bc}^{\prime \,+}$ & \multicolumn{1}{r}{$1.093$} & \multicolumn{1}{r}{$%
1.71$} & \multicolumn{1}{r}{$1.525$} & \multicolumn{1}{r}{$1.52$} & --- & 
\multicolumn{1}{r}{$1.990$} & --- & --- \\[2pt] 
$\Omega _{bc}^{0}$ & \multicolumn{1}{r}{$0.034$} & \multicolumn{1}{r}{$0.08$}
& \multicolumn{1}{r}{$-0.119$} & \multicolumn{1}{r}{$0.45$} & 
\multicolumn{1}{r}{$0.009$} & \multicolumn{1}{r}{$0.368$} & 
\multicolumn{1}{r}{$0.397$} & --- \\[2pt] 
$\Omega _{bc}^{\prime \,0}$ & \multicolumn{1}{r}{$-0.106$} & 
\multicolumn{1}{r}{$-0.27$} & --- & \multicolumn{1}{r}{$-0.61$} & --- & 
\multicolumn{1}{r}{$-0.542$} & --- & --- \\[2pt] 
$\Omega _{bcc}^{+}$ & \multicolumn{1}{r}{$0.505$} & \multicolumn{1}{r}{$0.54$%
} & \multicolumn{1}{r}{$0.476$} & \multicolumn{1}{r}{$0.53$} & 
\multicolumn{1}{r}{$0.475$} & --- & \multicolumn{1}{r}{$0.502$} & --- \\[2pt] 
$\Xi _{bb}^{-}$ & \multicolumn{1}{r}{$0.086$} & \multicolumn{1}{r}{$0.23$} & 
\multicolumn{1}{r}{$0.236$} & \multicolumn{1}{r}{$0.18$} & 
\multicolumn{1}{r}{$0.251$} & \multicolumn{1}{r}{$0.251$} & 
\multicolumn{1}{r}{$0.190$} & --- \\[2pt] 
$\Xi _{bb}^{0}$ & \multicolumn{1}{r}{$-0.432$} & \multicolumn{1}{r}{$-0.70$}
& \multicolumn{1}{r}{$-0.722$} & \multicolumn{1}{r}{$-0.53$} & 
\multicolumn{1}{r}{$-0.742$} & \multicolumn{1}{r}{$-0.742$} & 
\multicolumn{1}{r}{$-0.657$} & --- \\[2pt] 
$\Omega _{bb}^{-}$ & \multicolumn{1}{r}{$0.043$} & \multicolumn{1}{r}{$0.12$}
& \multicolumn{1}{r}{$0.100$} & \multicolumn{1}{r}{$0.04$} & 
\multicolumn{1}{r}{$0.101$} & \multicolumn{1}{r}{$0.101$} & 
\multicolumn{1}{r}{$0.109$} & --- \\[2pt] 
$\Omega _{bbc}^{0}$ & \multicolumn{1}{r}{$-0.205$} & \multicolumn{1}{r}{$%
-0.21$} & \multicolumn{1}{r}{$-0.197$} & \multicolumn{1}{r}{$-0.20$} & 
\multicolumn{1}{r}{$-0.193$} & --- & \multicolumn{1}{r}{$-0.203$} & --- \\[2pt] 
$\left| \Sigma _{b}^{0}\rightarrow \Lambda _{b}^{0}\right| $ & 
\multicolumn{1}{r}{$1.052$} & \multicolumn{1}{r}{$1.61$} & --- & --- & --- & 
--- & --- & --- \\[2pt] 
$\left| \Xi _{b}^{\prime \,-}\rightarrow \Xi _{b}^{-}\right| $ & 
\multicolumn{1}{r}{$0.082$} & \multicolumn{1}{r}{$0.16$} & --- & --- & --- & 
--- & --- & --- \\[2pt] 
$\left| \Xi _{b}^{\prime \,0}\rightarrow \Xi _{b}^{0}\right| $ & 
\multicolumn{1}{r}{$0.917$} & \multicolumn{1}{r}{$1.41$} & --- & --- & --- & 
--- & --- & --- \\[2pt] 
$\left| \Xi _{bc}^{\prime \,0}\rightarrow \Xi _{bc}^{0}\right| $ & 
\multicolumn{1}{r}{$0.508$} & \multicolumn{1}{r}{$0.70$} & --- & --- & --- & 
--- & --- & --- \\[2pt] 
$\left| \Xi _{bc}^{\prime \,+}\rightarrow \Xi _{bc}^{+}\right| $ & 
\multicolumn{1}{r}{$0.277$} & \multicolumn{1}{r}{$0.62$} & --- & --- & --- & 
--- & --- & --- \\[2pt] 
$\left| \Omega _{bc}^{\prime \,0}\rightarrow \Omega _{bc}^{0}\right| $ & 
\multicolumn{1}{r}{$0.443$} & \multicolumn{1}{r}{$0.56$} & --- & --- & --- & 
--- & --- & --- \\ \hline
\multicolumn{9}{l}{$^{**}$~Primes on states of $\Xi _{bc}$ and $\Omega _{bc}$ different from \cite{FGIK06} (opposite), as explained in the text.}
\end{tabular}
\end{table}%

\begin{table}[t] \centering%
\caption{Magnetic moments (in nuclear magnetons) of $J=\frac{3}{2}$ bottom
baryons calculated in the bag model (Bag) and in other approaches as
described in the text.\label{t4.5}} 
\begin{tabular}{lccccc}
\hline
Baryons & Bag & Nonrel & \cite{S96,AHNV07} & \cite{PRV08,PMV09} & \cite
{AAO09} \\ \hline
$\Sigma _{b}^{*\,-}$ & \multicolumn{1}{r}{$-1.271$} & \multicolumn{1}{r}{$%
-1.92$} & --- & \multicolumn{1}{r}{$-1.657$} & \multicolumn{1}{r}{$-1.50\pm
0.36$} \\[2pt] 
$\Sigma _{b}^{*\,0}$ & \multicolumn{1}{r}{$0.537$} & \multicolumn{1}{r}{$%
0.87 $} & --- & \multicolumn{1}{r}{$0.792$} & \multicolumn{1}{r}{$0.50\pm
0.15$} \\[2pt] 
$\Sigma _{b}^{*\,+}$ & \multicolumn{1}{r}{$2.346$} & \multicolumn{1}{r}{$%
3.56 $} & --- & \multicolumn{1}{r}{$3.239$} & \multicolumn{1}{r}{$2.52\pm
0.50$} \\[2pt] 
$\Xi _{b}^{*\,-}$ & \multicolumn{1}{r}{$-1.088$} & \multicolumn{1}{r}{$-1.60$%
} & --- & \multicolumn{1}{r}{$-1.098$} & \multicolumn{1}{r}{$-1.42\pm 0.35$}
\\[2pt] 
$\Xi _{b}^{*\,0}$ & \multicolumn{1}{r}{$0.690$} & \multicolumn{1}{r}{$1.19$}
& --- & \multicolumn{1}{r}{$1.042$} & \multicolumn{1}{r}{$0.50\pm 0.15$} \\[2pt] 
$\Omega _{b}^{*\,-}$ & \multicolumn{1}{r}{$-0.919$} & \multicolumn{1}{r}{$%
-1.28$} & --- & \multicolumn{1}{r}{$-1.201$} & \multicolumn{1}{r}{$-1.40\pm
0.35$} \\[2pt] 
$\Xi _{bc}^{*\,0}$ & \multicolumn{1}{r}{$-0.257$} & \multicolumn{1}{r}{$%
-0.60 $} & \multicolumn{1}{r}{$-0.712$} & \multicolumn{1}{r}{$-0.568$} & ---
\\[2pt] 
$\Xi _{bc}^{*\,+}$ & \multicolumn{1}{r}{$1.414$} & \multicolumn{1}{r}{$2.19$}
& \multicolumn{1}{r}{$2.270$} & \multicolumn{1}{r}{$2.052$} & --- \\[2pt] 
$\Omega _{bc}^{*\,0}$ & \multicolumn{1}{r}{$-0.111$} & \multicolumn{1}{r}{$%
-0.28$} & \multicolumn{1}{r}{$-0.261$} & \multicolumn{1}{r}{$-0.317$} & ---
\\[2pt] 
$\Omega _{bcc}^{*\,+}$ & \multicolumn{1}{r}{$0.659$} & \multicolumn{1}{r}{$%
0.72$} & --- & \multicolumn{1}{r}{$0.651$} & --- \\[2pt] 
$\Xi _{bb}^{*\,-}$ & \multicolumn{1}{r}{$-0.652$} & \multicolumn{1}{r}{$%
-1.05 $} & \multicolumn{1}{r}{$-1.110$} & \multicolumn{1}{r}{$-0.952$} & ---
\\[2pt] 
$\Xi _{bb}^{*\,0}$ & \multicolumn{1}{r}{$0.916$} & \multicolumn{1}{r}{$1.74$}
& \multicolumn{1}{r}{$1.870$} & \multicolumn{1}{r}{$1.577$} & --- \\[2pt] 
$\Omega _{bb}^{*\,-}$ & \multicolumn{1}{r}{$-0.522$} & \multicolumn{1}{r}{$%
-0.73$} & \multicolumn{1}{r}{$-0.662$} & \multicolumn{1}{r}{$0.711$} & ---
\\[2pt] 
$\Omega _{bbc}^{*\,0}$ & \multicolumn{1}{r}{$0.225$} & \multicolumn{1}{r}{$%
0.27$} & --- & \multicolumn{1}{r}{$0.216$} & --- \\[2pt] 
$\Omega _{bbb}^{-}$ & \multicolumn{1}{r}{$-0.194$} & \multicolumn{1}{r}{$%
-0.18$} & \multicolumn{1}{r}{$-0.180$} & \multicolumn{1}{r}{$-0.195$} & ---
\\ \hline
\end{tabular}
\end{table}%

Care must be taken when one tries to compare various expressions and results
of the calculations with earlier works because of some mess-up in the
notations of primed and unprimed states for the single heavy baryons $\Xi
_{Q}$ and $\Xi _{Q}^{\prime }$. Usually \cite{KR10} the physical $\Xi _{Q}$
state is assumed to be that which contains a pair of light quarks mostly in $%
S=0$ (antisymmetric) state $\left| \Xi _{Q}\right\rangle \sim \left|
[q_{1}q_{2}]Q\right\rangle $ where $q_{i}$ denotes the light and $Q$ the heavy
quarks. Respectively the primed state $\Xi _{Q}^{\prime }$ is mostly $S=1$
(symmetric) state $\left| \Xi _{Q}^{\prime }\right\rangle \sim \left|
\{q_{1}q_{2}\}Q\right\rangle $. Often the notations $\Xi _{Q}$ ($\Xi 
_{Q}^{\prime }$)$\,$ are simply used to denote pure antisymmetric
(symmetric) states. When the quark model was young the opposite convention
was in use. Such old fashioned (opposite) convention has been used for
designating the primed states in Refs.~\cite{FLNC81,JS77,BD83}. This
circumstance must be taken into account when comparing their results with
ours.

Some complications arise when we want to compare our predictions for doubly
heavy baryons $\Xi _{bc},\Xi _{bc}^{\prime }$ and $\Omega _{bc},\Omega
_{bc}^{\prime }$ with the unmixed results obtained using quark ordering
scheme representing heavy diquark picture \cite{FGIK06,AHNV07,PRV08,FGIK09},
in which the spins of the two heaviest quarks are coupled to form symmetric $%
\{Q_{1}Q_{2}\}$ or antisymmetric $[Q_{1}Q_{2}]$ diquark. At first sight such
scheme seems to follow the recipe of Ref.~\cite{FLNC81} that the two quarks
closest in mass must be (anti)symmetrized as the first two. But the fact that the two
quarks are the heaviest does not mean that they are the closest in mass. With
respect to the colour-hyperfine interaction $u$ (or $d$) and $c$ quarks are
closer than $c$ and $b$ (see Table~\ref{t3.2}). Meanwhile, in the
case of two identical heavy quarks the heavy diquark picture is a perfect
choice. Of course, we can compare their predictions with our unmixed results
corresponding to the quark ordering $(Q_{1}Q_{2})q$ and obtain good
qualitative agreement. However, our unmixed states are not physical states,
therefore such comparison between presumably physical and nonphysical
states seems to be unsatisfactory. On the other hand, we see from Table~%
\ref{t3.2} that our unprimed state is predominantly the symmetric heavy
diquark state with some (not very small) admixture of antisymmetric state,
i.e. $\left| B\right\rangle =C_{1}\left| [Q_{1}Q_{2}]q_{3}\right\rangle
+C_{2}\left| \{Q_{1}Q_{2}\}q_{3}\right\rangle $, where $C_{2}^{2}>C_{1}^{2}$%
. So, it makes some sense to denote the symmetric heavy diquark state as $%
\left| B\right\rangle $ and antisymmetric one as $\left| B^{\prime
}\right\rangle $. Such convention has been chosen in Refs.~\cite
{AHNV07,PRV08,FGIK09}. However, in Ref.~\cite{FGIK06} the opposite
convention has been used. For convenience and in order to have a more
consistent representation we have renamed (in Table~\ref{t4.4}) their $%
\Xi _{bc}$ and $\Omega _{bc}$ states as $\Xi _{bc}^{\prime }$ and $\Omega
_{bc}^{\prime }$ (and correspondingly $\Xi _{bc}^{\prime }$, $\Omega
_{bc}^{\prime }$ as $\Xi _{bc}$, $\Omega _{bc}$).

Now let us focus on the results presented in Tables~\ref
{t4.1}--\ref{t4.5}. The first impression is that almost all collected
predictions (with only several exceptions) as a whole give us a relatively
consistent picture. This is the consequence of underlying symmetry shared by
the models. But can we understand the differences? Not always, but sometimes
we can. As a first step it is not a bad idea to compare our bag model
predictions with the results given by simple nonrelativistic model (columns
3 of the Tables~\ref{t4.1}, \ref{t4.3}--\ref{t4.5}). We see that the 
qualitative picture in both variants is similar, while numerical values
differ, sometimes substantially. Maybe the most intriguing feature is the
opposite sign of predicted magnetic moments in the bag and nonrelativistic
models for $\Xi _{cc}^{++}$ and $\Xi _{cc}^{*\,+}$ baryons. The inspection
of predictions presented in Tables~\ref{t4.1}--\ref{t4.3} shows that
there is now common agreement on the signs of these magnetic moments. The
explanation of such disagreement is quite simple. The magnetic moment of $%
\Xi _{cc}^{++}$ is given by the expression $\mu (\Xi _{cc}^{++})=\frac{2}{9}%
(4\bar{\mu}_{c}-\bar{\mu}_{u})$ (see Table~\ref{tA.1}). The sign of this
magnetic moment depends on what is ``stronger'' -- two $c$ quarks or one $u$
quark. In the nonrelativistic model the $u$ quark overcomes the $c$-duet,
while in our bag model the $c$ quarks defeat a single $u$ quark. The magnetic
moment of $\Xi _{cc}^{*\,+}$ is related to $\mu (\Xi _{cc}^{++})$ by Eq.~(%
\ref{eq2.24a}), therefore its sign must be the same.

Contrary to the naive nonrelativistic model, in the bag model the magnetic
moment of a quark depends on baryon, the owner of this quark. There are two
effects that make the quark magnetic moments in the case of heavy baryons
smaller. Firstly, the dependence of the quark magnetic moment on the bag
radius. A heavier baryon has a smaller bag radius and this leads to a smaller
quark magnetic moment in accordance. The second reason is the relative
strength of the c.m.m. corrections. In the case of light hadrons these
corrections are stronger and therefore lead to larger magnetic moments than
in the case of heavy baryons. For example, in the proton the value of the
c.m.m. corrected magnetic moment of $u$ quark is $\mu _{u}(P)=1.924~\mu _{N}$%
, while in the $\Xi _{b}$ baryon it is only $\mu _{u}(\Xi _{b})=1.168~\mu
_{N}$. For heavier baryon $\Xi _{bb}$ it becomes even smaller, $\mu _{u}(\Xi
_{bb})=1.036~\mu _{N}$. The similar feature of the magnetic moment of the strange
quark also cannot be ignored. For example, $\mu _{s}(\Omega
_{b})=-0.425~\mu _{N}$, while $\mu _{s}(\Omega _{bb})=-0.390~\mu _{N}$. It
is this difference that was at the root of the failure of the quark model
relation $3[4\Omega _{bb}^{-}+\Omega _{b}^{-}]=5[2\Omega _{bb}^{*\,-}-\Omega
_{b}^{*\,-}]$ in Sec.~3. For charmed quarks the dependence diminishes, but 
remains appreciable. For example, $\mu _{c}(\Lambda _{c}^{+})=0.411~\mu _{N}$%
, while in the triply heavy baryon $\mu _{c}(\Omega _{ccc}^{++})=0.379~\mu
_{N}$. In contrast, the magnetic moments of bottom quarks are almost
insensible to the baryon they live in ($\mu _{b}(\Lambda
_{b}^{0})=-0.066~\mu _{N}$ and $\mu _{b}(\Omega _{bbb}^{-})=-0.065~\mu _{N}$%
, for example). There are several baryons (e.g., $\Lambda _{c}$, $\Omega
_{ccc}$, $\Lambda _{b}$, $\Omega _{bcc}$, $\Omega _{bcc}^{*}$, $\Omega
_{bbc} $, $\Omega _{bbc}^{*}$, $\Omega _{bbb}$) the magnetic moments of which
depend on the magnetic moments of the heavy quarks only. We expect that in
these cases the results obtained in the bag model and in the nonrelativistic
one must be similar. The differences indeed do not exceed $10\%$. In all
other cases the bag model predicted values of magnetic moments are smaller
than corresponding nonrelativistic results (sometimes distinctly).

The magnetic moments of heavy baryons are unlikely to be measured in the
nearest future. In such situation any indirect estimate of these quantities
could be helpful. Some useful information can be extracted from the mass
spectra of baryons. Magnetic moments of quarks are proportional to the
chromomagnetic moments which determine the colour-hyperfine splitting of
baryon masses. Using this fact the magnetic moments of $\Lambda _{c}$ and $%
\Lambda _{b}$ can be obtained \cite{KL06}. The predictions are $\mu
_{c}(\Lambda _{c}^{+})=0.43~\mu _{N}$ and $\mu _{b}(\Lambda
_{b}^{0})=-0.067~\mu _{N}$, in excellent coincidence with our results. These
values are also consistent with almost all other predictions with the
exception of spectral sum rules \cite{ZHY97}, where the source of deviation
is the contribution from higher-dimension condensates.

We think it could be reasonable to continue the comparison of bag model
predictions with the results obtained in other approaches with the simplest
case of $J=\frac{3}{2}$ bottom baryons (Table~\ref{t4.5}). We see
immediately that our predictions are compatible with the light cone sum
rules \cite{AAO09}, while the agreement between nonrelativistic model
predictions and light cone sum rules is not so good. The results obtained
using hypercentral model \cite{PRV08,PMV09} are, as a rule, somewhere
between naive nonrelativistic predictions and ours. For example, their
predictions for the magnetic moments of triply heavy baryons agree well with
our results, while in other cases they are closer to the predictions of
simple nonrelativistic model. Variational calculations \cite{AHNV07} do not
differ substantially from the results obtained using naive nonrelativistic
model.

The situation is similar for the $J=\frac{3}{2}$ charmed baryons (Table~\ref
{t4.3}). Bag model predictions for $\Sigma _{c}^{*\,0}$, $\Xi _{c}^{*\,0}$, $%
\Xi _{c}^{*\,+}$, and $\Omega _{c}^{*\,0}$ are compatible with the light
cone sum rules again, while in the case of $\Sigma _{c}^{*\,+}$and $\Sigma
_{c}^{*\,++}$ our values are somewhat lower than the light cone results. The
hypercentral predictions on the average are closer to naive nonrelativistic
results, but these for $\Sigma _{c}^{*\,0}$, $\Xi _{c}^{*\,0}$ are closer to
ours. The chiral model \cite{SDCG10} predicts larger values in all cases,
even larger than nonrelativistic results. Variational results are close to
naive nonrelativistic predictions in this case again.

Before proceeding with the case of $J=\frac{1}{2}$ baryons let us revert for
a moment to the comparison of our bag model predictions with the predictions
obtained in the simple nonrelativistic approach. There exists one more
correspondence between bag model results and the nonrelativistic
predictions. In both cases for the baryons containing three differently
flavoured quarks the state mixing effect caused by the colour-hyperfine
interaction was taken into account. This is a significant improvement which
can lead to substantial shifts of the predicted magnetic moments. For the
mixed magnetic moments we have reasonable qualitative agreement between
predictions obtained in both models. Another approach in which the state
mixing is taken into account (by definition) is the Faddeev formalism \cite
{S96}. The predictions obtained using this method do not differ very much
from naive nonrelativistic results. To see what is the importance of the
state mixing effect one can compare the result of Ref.~\cite{S96} for
baryons $\Xi _{bc}^{0}$, $\Xi _{bc}^{+}$, and $\Omega _{bc}^{0}$ with the
corresponding results obtained in Ref.~\cite{AHNV07} where this mixing was
ignored. Note that predicted magnetic moments can differ by almost 40 times
(in extreme case of $\Omega _{bc}^{0}$). The importance of this effect for
the models with effective colour-hyperfine interaction has been known for
years \cite{FLNC81}, nevertheless, in many calculations it was
systematically ignored for various reasons (peculiarities of the model,
technical difficulties, etc.) This seems to be the case for all other
calculations we are going to compare our predictions with. A special
exception is pure chiral models in which the state mixing of this type is
naturally absent because of different type of effective interaction used.
This is the reason of some qualitative difference between predictions for
magnetic moments of heavy baryons obtained in chiral approach and in models
based on the effective colour-hyperfine interaction. Therefore when we
compare our results for baryons $\Xi _{c}^{+},\Xi _{c}^{\prime \,+}$; $\Xi
_{b}^{0},\Xi _{b}^{\prime \,0}$; $\Xi _{bc}^{0},\Xi _{bc}^{\prime \,0}$; $%
\Xi _{bc}^{+},\Xi _{bc}^{\prime \,+}$; and $\Omega _{bc}^{0},\Omega
_{bc}^{\prime \,0}$ with predictions of others we must keep all theses
circumstances in mind.

With these not very short preliminaries we can now proceed
comparing of our predictions with other results. We see that all models, as
expected, give very similar predictions in the case of triply heavy $J=\frac{%
1}{2}$ baryons $\Omega _{bcc}$ and $\Omega _{bbc}$. The agreement of bag
model results with the sum rules in the case of $J=\frac{1}{2}$ baryons is
not so good. For example, for $\Xi _{c}^{0}$ and $\Xi _{c}^{+}$ baryons the
light cone sum rules predict the values similar to results obtained in
chiral model \cite{SDCG10} and in chiral perturbation theory \cite{S94}. Our
predictions in these cases differ substantially. The state mixing effect
acts in opposite direction as chiral corrections. Could these two effects if
applied simultaneously compensate each other? In any case it should depend
on the model. Possibly it could happen in models with some mixture of
one-gluon-exchange and Goldstone-boson-exchange induced interactions.
Hypercentral predictions in the charm sector are again somewhere between
simple nonrelativistic results and ours (some closer to ours, some to naive
nonrelativistic), but for the bottom baryons they are closer to the
predictions obtained in simple nonrelativistic approach. Almost all other
predictions are closer to nonrelativistic results and provide larger values
than ours. For example, the predictions of Ref.~\cite{JR04} for unmixed
moments are almost indistinguishable from the naive nonrelativistic
predictions. However, we found out, with some surprise, that our predictions
for magnetic moments of $J=\frac{1}{2}$ baryons (at least unmixed) resemble
the results obtained in Ref.~\cite{FGIK06} including positive sign of $\mu
(\Xi _{cc}^{++})$. The models are very different, and we have no reasonable
explanation of this resemblance. Could the reason be a common relativistic
nature? Initially both them were formulated as relativistic nonlocal field
theories. It may be, but the other two relativistic models \cite{JR04,BD83}
behave very much like their nonrelativistic neighbours.

In summary, we have used the improved bag model to calculate magnetic
moments of $J=\frac{1}{2}$ and $J=\frac{3}{2}$ heavy baryons without
introducing any new parameters and obtained encouraging results. The
status of quark model relations connecting magnetic moments of various
baryons was revisited. A part of work was devoted to study the state mixing
induced by the colour-hyperfine interaction. It has been shown that for the
baryons $\Xi _{c}^{+},\Xi _{c}^{\prime \,+}$; $\Xi _{b}^{0},\Xi _{b}^{\prime
\,0}$; $\Xi _{bc}^{0},\Xi _{bc}^{\prime \,0}$; $\Xi _{bc}^{+},\Xi
_{bc}^{\prime \,+}$; and $\Omega _{bc}^{0},\Omega _{bc}^{\prime \,0}$ this
mixing leads to appreciable shifts of the unmixed magnetic moments and, in
order to have a consistent description, must be taken into account.

\pagebreak

\appendix 

\section{Appendix: explicit expressions for the magnetic moments of heavy
baryons}

In the three tables below we present expressions for the magnetic moments of 
$\mathrm{spin\,}\frac{1}{2}$ and $\mathrm{spin\,}\frac{3}{2}$ charmed and
bottom baryons. For simplicity the shorthand notations ($\mu _{q}\rightarrow
q$, $\bar{\mu}_{q}\rightarrow \bar{q}$) are used. The entries in columns 4
were obtained by setting $(u,c)=\frac{2}{3}(\bar{u},\bar{c})$, $(d,s,b)=-%
\frac{1}{3}(\bar{d},\bar{s},\bar{b})$, and assuming isospin symmetry ($\bar{u%
}=\bar{d}$).

In the case of baryons containing three quarks of different flavours the
quark arrangement with the heaviest quark placed as the third one in the
spin coupling scheme has been chosen. The entries of the column~3 for other
arrangements can be easily obtained by simple quark renaming.

\begin{table}[!b] \centering%
\caption{Composition of $J=\frac{1}{2}$ charmed baryon magnetic moments in terms of magnetic
moments of individual quarks (column 3) and corresponding reduced quantities
(column 4).\label{tA.1}} 
\begin{tabular}{cccc}
\hline
Particles & quark ordering & $\mu _{\mathrm{B}}^{0}$ & $\mu _{\mathrm{B}%
}^{0} $ \\ \hline \\[-8pt]
$\Lambda _{c}^{+}$ & $[ud]c$ & $c$ & $\frac{2}{3}\bar{c}$ \\[4pt] 
$\Sigma _{c}^{+}$ & $\{ud\}c$ & $\frac{1}{3}(2u+2d-c)$ & $\frac{2}{9}(\bar{u}%
-\bar{c})$ \\[4pt] 
$\Sigma _{c}^{+}\rightarrow \Lambda _{c}^{+}$ & $\{ud\}c\rightarrow [ud]c$ & 
$\frac{1}{\sqrt{3}}(d-u)$ & $-\frac{1}{\sqrt{3}}\bar{u}$ \\[4pt] 
$\Sigma _{c}^{0}$ & $ddc$ & $\frac{1}{3}(4d-c)$ & $-\frac{2}{9}(2\bar{u}+%
\bar{c})$ \\[4pt] 
$\Sigma _{c}^{++}$ & $uuc$ & $\frac{1}{3}(4u-c)$ & $\frac{2}{9}(4\bar{u}-%
\bar{c})$ \\[4pt] 
$\Xi _{c}^{0},\Xi _{c}^{\prime \,0}$ & $[ds]c$ & $c$ & $\frac{2}{3}\bar{c}$
\\[4pt] 
\textquotedbl & $\{ds\}c$ & $\frac{1}{3}(2d+2s-c)$ & $-\frac{2}{9}(\bar{u}+\bar{s}+%
\bar{c})$ \\[4pt] 
\textquotedbl & $\{ds\}c\rightarrow [ds]c$ & $\frac{1}{\sqrt{3}}(s-d)$ & $\frac{1}{3%
\sqrt{3}}(\bar{u}-\bar{s})$ \\[4pt] 
$\Xi _{c}^{+},\Xi _{c}^{\prime \,+}$ & $[us]c$ & $c$ & $\frac{2}{3}\bar{c}$
\\[4pt] 
\textquotedbl & $\{us\}c$ & $\frac{1}{3}(2u+2s-c)$ & $\frac{2}{9}(2\bar{u}-\bar{s}-%
\bar{c})$ \\[4pt] 
\textquotedbl & $\{us\}c\rightarrow [us]c$ & $\frac{1}{\sqrt{3}}(s-u)$ & $-\frac{1}{3%
\sqrt{3}}(2\bar{u}+\bar{s})$ \\[4pt] 
$\Omega _{c}^{0}$ & $ssc$ & $\frac{1}{3}(4s-c)$ & $-\frac{2}{9}(2\bar{s}+%
\bar{c})$ \\[4pt] 
$\Xi _{cc}^{+}$ & $ccd$ & $\frac{1}{3}(4c-d)$ & $\frac{1}{9}(\bar{u}+8\bar{c}%
)$ \\[4pt] 
$\Xi _{cc}^{++}$ & $ccu$ & $\frac{1}{3}(4c-u)$ & $-\frac{2}{9}(\bar{u}-4\bar{%
c})$ \\[4pt] 
$\Omega _{cc}^{+}$ & $ccs$ & $\frac{1}{3}(4c-s)$ & $\frac{1}{9}(\bar{s}+8%
\bar{c})$ \\[4pt] \hline
\end{tabular}
\end{table}%

\begin{table}
\centering%
\caption{Composition of $J=\frac{1}{2}$ bottom baryon magnetic moments in terms of magnetic
moments of individual quarks (column 3) and corresponding reduced quantities
(column 4).\label{tA.2}} 
\begin{tabular}{lccc}
\hline
Particles & quark ordering & $\mu _{\mathrm{B}}^{0}$ & $\mu _{\mathrm{B}%
}^{0} $ \\ \hline \\[-8pt]
\multicolumn{1}{c}{$\Lambda _{b}^{0}$} & $[ud]b$ & $b$ & $-\frac{1}{3}\bar{b}
$ \\[4pt] 
\multicolumn{1}{c}{$\Sigma _{b}^{0}$} & $\{ud\}b$ & $\frac{1}{3}(2u+2d-b)$ & 
$\frac{1}{9}(2\bar{u}+\bar{b})$ \\[4pt] 
\multicolumn{1}{c}{$\Sigma _{b}^{0}\rightarrow \Lambda _{b}^{0}$} & $%
\{ud\}b\rightarrow [ud]b$ & $\frac{1}{\sqrt{3}}(d-u)$ & $-\frac{1}{\sqrt{3}}%
\bar{u}$ \\[4pt] 
\multicolumn{1}{c}{$\Sigma _{b}^{-}$} & $ddb$ & $\frac{1}{3}(4d-b)$ & $-%
\frac{1}{9}(4\bar{u}-\bar{b})$ \\[4pt] 
\multicolumn{1}{c}{$\Sigma _{b}^{+}$} & $uub$ & $\frac{1}{3}(4u-b)$ & $\frac{%
1}{9}(8\bar{u}+\bar{b})$ \\[4pt] 
\multicolumn{1}{c}{$\Xi _{b}^{-},\Xi _{b}^{\prime \,-}$} & $[ds]b$ & $b$ & $-%
\frac{1}{3}\bar{b}$ \\[4pt] 
\multicolumn{1}{c}{\textquotedbl} & $\{ds\}b$ & $\frac{1}{3}(2d+2s-b)$ & $-\frac{1}{9}%
(2\bar{u}+2\bar{s}-\bar{b})$ \\[4pt] 
\multicolumn{1}{c}{\textquotedbl} & $\{ds\}b\rightarrow [ds]b$ & $\frac{1}{\sqrt{3}}%
(s-d)$ & $\frac{1}{3\sqrt{3}}(\bar{u}-\bar{s})$ \\[4pt] 
\multicolumn{1}{c}{$\Xi _{b}^{0},\Xi _{b}^{\prime \,0}$} & $[us]b$ & $b$ & $-%
\frac{1}{3}\bar{b}$ \\[4pt] 
\multicolumn{1}{c}{\textquotedbl} & $\{us\}b$ & $\frac{1}{3}(2u+2s-b)$ & $\frac{1}{9}%
(4\bar{u}-2\bar{s}+\bar{b})$ \\[4pt] 
\multicolumn{1}{c}{\textquotedbl} & $\{us\}b\rightarrow [us]b$ & $\frac{1}{\sqrt{3}}%
(s-u)$ & $-\frac{1}{3\sqrt{3}}(2\bar{u}+\bar{s})$ \\[4pt] 
\multicolumn{1}{c}{$\Omega _{b}^{-}$} & $ssb$ & $\frac{1}{3}(4s-b)$ & $-%
\frac{1}{9}(4\bar{s}-\bar{b})$ \\[4pt] 
\multicolumn{1}{c}{$\Xi _{bc}^{0},\Xi _{bc}^{\prime \,0}$} & $[dc]b$ & $b$ & 
$-\frac{1}{3}\bar{b}$ \\[4pt] 
\multicolumn{1}{c}{\textquotedbl} & $\{dc\}b$ & $\frac{1}{3}(2d+2c-b)$ & $-\frac{1}{9}%
(2\bar{u}-4\bar{c}-\bar{b})$ \\[4pt] 
\multicolumn{1}{c}{\textquotedbl} & $\{dc\}b\rightarrow [dc]b$ & $\frac{1}{\sqrt{3}}%
(c-d)$ & $\frac{1}{3\sqrt{3}}(\bar{u}+2\bar{c})$ \\[4pt] 
\multicolumn{1}{c}{$\Xi _{bc}^{+},\Xi _{bc}^{\prime \,+}$} & $[uc]b$ & $b$ & 
$-\frac{1}{3}\bar{b}$ \\[4pt] 
\multicolumn{1}{c}{\textquotedbl} & $\{uc\}b$ & $\frac{1}{3}(2u+2c-b)$ & $\frac{1}{9}%
(4\bar{u}+4\bar{c}+\bar{b})$ \\[4pt] 
\multicolumn{1}{c}{\textquotedbl} & $\{uc\}b\rightarrow [uc]b$ & $\frac{1}{\sqrt{3}}%
(c-u)$ & $\frac{2}{3\sqrt{3}}(\bar{u}-\bar{c})$ \\[4pt] 
\multicolumn{1}{c}{$\Omega _{bc}^{0},\Omega _{bc}^{\prime \,0}$} & $[sc]b$ & 
$b$ & $-\frac{1}{3}\bar{b}$ \\[4pt] 
\multicolumn{1}{c}{\textquotedbl} & $\{sc\}b$ & $\frac{1}{3}(2s+2c-b)$ & $-\frac{1}{9}%
(2\bar{s}-4\bar{c}-\bar{b})$ \\[4pt] 
\multicolumn{1}{c}{\textquotedbl} & $\{sc\}b\rightarrow [sc]b$ & $\frac{1}{\sqrt{3}}%
(c-s)$ & $\frac{1}{3\sqrt{3}}(\bar{s}+2\bar{c})$ \\[4pt] 
\multicolumn{1}{c}{$\Omega _{bcc}^{+}$} & $ccb$ & $\frac{1}{3}(4c-b)$ & $%
\frac{1}{9}(8\bar{c}-\bar{b})$ \\[4pt] 
\multicolumn{1}{c}{$\Xi _{bb}^{-}$} & $bbd$ & $\frac{1}{3}(4b-d)$ & $\frac{1%
}{9}(\bar{u}-4\bar{b})$ \\[4pt] 
\multicolumn{1}{c}{$\Xi _{bb}^{0}$} & $bbu$ & $\frac{1}{3}(4b-u)$ & $-\frac{2%
}{9}(\bar{u}+2\bar{b})$ \\[4pt] 
\multicolumn{1}{c}{$\Omega _{bb}^{-}$} & $bbs$ & $\frac{1}{3}(4b-s)$ & $%
\frac{1}{9}(\bar{s}-4\bar{b})$ \\[4pt] 
\multicolumn{1}{c}{$\Omega _{bbc}^{0}$} & $bbc$ & $\frac{1}{3}(4b-c)$ & $-%
\frac{2}{9}(\bar{c}+2\bar{b})$ \\[4pt] \hline
\end{tabular}
\end{table}%

\begin{table}
\centering%
\caption{Composition of $J=\frac{3}{2}$ charmed and bottom baryon magnetic moments in terms of
magnetic moments of individual quarks (column 3) and corresponding reduced
quantities (column 4).\label{tA.3}} 
\begin{tabular}{lccc}
\hline
Particles & quark content & $\mu _{\mathrm{B}}^{0}$ & $\mu _{\mathrm{B}}^{0}$
\\ \hline \\[-8pt]
$\Sigma _{c}^{*\,0}$ & $ddc$ & $2d+c$ & $-\frac{2}{3}(\bar{u}-\bar{c})$ \\[4pt] 
$\Sigma _{c}^{*\,+}$ & $\{ud\}c$ & $u+d+c$ & $\frac{1}{3}(\bar{u}+2\bar{c})$
\\[4pt] 
$\Sigma _{c}^{*\,++}$ & $uuc$ & $2u+c$ & $\frac{2}{3}(2\bar{u}+\bar{c})$ \\[4pt] 
$\Xi _{c}^{*\,0}$ & $\{ds\}c$ & $d+s+c$ & $\frac{1}{3}(2\bar{c}-\bar{u}-\bar{%
s})$ \\[4pt] 
$\Xi _{c}^{*\,+}$ & $\{us\}c$ & $u+s+c$ & $\frac{1}{3}(2\bar{c}+2\bar{u}-%
\bar{s})$ \\[4pt] 
$\Omega _{c}^{*\,0}$ & $ssc$ & $2s+c$ & $-\frac{2}{3}(\bar{s}-\bar{c})$ \\[4pt] 
$\Xi _{cc}^{*\,+}$ & $ccd$ & $2c+d$ & $-\frac{1}{3}(\bar{u}-4\bar{c})$ \\[4pt] 
$\Xi _{cc}^{*\,++}$ & $ccu$ & $2c+u$ & $\frac{2}{3}(\bar{u}+2\bar{c})$ \\[4pt] 
$\Omega _{cc}^{*\,+}$ & $ccs$ & $2c+s$ & $\frac{1}{3}(4\bar{c}-\bar{s})$ \\[4pt] 
$\Omega _{ccc}^{++}$ & $ccc$ & $3c$ & $2\bar{c}$ \\[4pt] 
$\Sigma _{b}^{*\,+}$ & $uub$ & $2u+b$ & $\frac{1}{3}(4\bar{u}-\bar{b})$ \\[4pt] 
$\Sigma _{b}^{*\,0}$ & $\{ud\}b$ & $u+d+b$ & $\frac{1}{3}(\bar{u}-\bar{b})$
\\[4pt] 
$\Sigma _{b}^{*\,-}$ & $ddb$ & $2d+b$ & $-\frac{1}{3}(2\bar{u}+\bar{b})$ \\[4pt] 
$\Xi _{b}^{*\,0}$ & $\{us\}b$ & $u+s+b$ & $\frac{1}{3}(2\bar{u}-\bar{s}-\bar{%
b})$ \\[4pt] 
$\Xi _{b}^{*\,-}$ & $\{ds\}b$ & $d+s+b$ & $-\frac{1}{3}(\bar{u}+\bar{s}+\bar{%
b})$ \\[4pt] 
$\Omega _{b}^{*\,-}$ & $ssb$ & $2s+b$ & $-\frac{1}{3}(2\bar{s}+\bar{b})$ \\[4pt] 
$\Xi _{bc}^{*\,+}$ & $\{uc\}b$ & $u+c+b$ & $\frac{1}{3}(2\bar{u}+2\bar{c}-%
\bar{b})$ \\[4pt] 
$\Xi _{bc}^{*\,o}$ & $\{dc\}b$ & $d+c+b$ & $-\frac{1}{3}(\bar{u}+\bar{b}-2%
\bar{c})$ \\[4pt] 
$\Omega _{bc}^{*\,0}$ & $\{sc\}b$ & $s+c+b$ & $-\frac{1}{3}(\bar{s}+\bar{b}-2%
\bar{c})$ \\[4pt] 
$\Omega _{bcc}^{*\,+}$ & $ccb$ & $2c+b$ & $\frac{1}{3}(4\bar{c}-\bar{b})$ \\[4pt] 
$\Xi _{bb}^{*\,0}$ & $bbu$ & $2b+u$ & $\frac{2}{3}(\bar{u}-\bar{b})$ \\[4pt] 
$\Xi _{bb}^{*\,-}$ & $bbd$ & $2b+d$ & $-\frac{1}{3}(\bar{u}+2\bar{b})$ \\[4pt] 
$\Omega _{bb}^{*\,-}$ & $bbs$ & $bb+s$ & $-\frac{1}{3}(\bar{s}+2\bar{b})$ \\[4pt] 
$\Omega _{bbc}^{*\,0}$ & $bbc$ & $2b+c$ & $\frac{2}{3}(\bar{c}-\bar{b})$ \\[4pt] 
$\Omega _{bbb}^{-}$ & $bbb$ & $3b$ & $-\bar{b}$ \\[4pt] \hline
\end{tabular}
\end{table}%

\pagebreak

\end{document}